%% file: paper_v10.tex
\documentclass[10pt,twocolumn]{IEEEtran2}

\usepackage{color}     
\usepackage{epsf,psfrag,amssymb,amsfonts,latexsym,cite,verbatim,enumerate,subfigure}
\usepackage{srcltx,amsmath,cases, graphicx}
\usepackage{times, multirow, array}
\usepackage[mathscr]{eucal}
\usepackage[justification=centering]{caption}

\input macros

\input labelfig.tex

\def\scalefig#1{\epsfxsize #1\textwidth}

\newcommand {\Ebb}{{\mathbb{E}}}

\def\beq{\begin{equation}}
\def\eeq{\end{equation}}
\def\beqa{\begin{eqnarray}}
\def\eeqa{\end{eqnarray}}
\def\beqa*{\begin{eqnarray*}}
\def\eeqa*{\end{eqnarray*}}

\newtheorem{theorem}{Theorem}

\newtheorem{remark}{Remark}

\newtheorem{algorithm}{Algorithm}
\newtheorem{problem}{Problem}

\title{\huge {Filter Design for Generalized Frequency-Division Multiplexing}}

\author{
 Seungyul Han, {\em Student~Member, IEEE}, Youngchul
Sung$^\dagger$\thanks{$^\dagger$Corresponding author}, {\em
Senior~Member, IEEE}, and Yong H. Lee, {\em Senior~Member, IEEE} \\
\thanks{The authors are with Dept. of Electrical Engineering,  KAIST, Daejeon 305-701, South
Korea. E-mail:\{sy.han, ycsung, and yohlee\}@kaist.ac.kr.  This
work was supported  by the Basic Science Research Program through
the National Research Foundation of Korea (NRF) funded by the
Ministry of Education (2013R1A1A2A10060852). A preliminary version of this work was submitted to GlobalSIP 2016 \cite{Han&Sung:16GlobalSIP}.}
}

\begin{document}

\maketitle

\begin{abstract}
In this paper, optimal filter design for generalized
frequency-division multiplexing (GFDM) is considered under two design criteria: rate maximization and  out-of-band (OOB) emission minimization.   First, the problem of GFDM
filter optimization  for rate maximization  is formulated by expressing the transmission rate of GFDM as a
function of GFDM filter coefficients. It is shown that Dirichlet filters are rate-optimal in  additive white Gaussian noise (AWGN)
channels with no carrier frequency offset (CFO) under linear
zero-forcing (ZF) or minimum mean-square error (MMSE) receivers,
but in general channels perturbed by CFO a properly designed nontrivial GFDM
filter can yield better performance than Dirichlet filters  by adjusting the
subcarrier waveform to cope with the channel-induced CFO. Next, the problem of GFDM filter design for OOB emission minimization is formulated by expressing the power spectral density (PSD) of the GFDM transmit signal as a function of GFDM filter coefficients, and it is shown that the OOB emission can be reduced significantly by designing the GFDM filter properly. Finally, joint design of GFDM filter and window for the two design criteria is considered.
\end{abstract}


\section{Introduction}

Recently active research is on-going regarding  waveform design
for 5G wireless communication beyond orthogonal frequency-division
multiplexing (OFDM) which has been used for
4G\cite{Wunder&etal:14ComMag}. The main drive for the design of a
new waveform is Internet-Of-Things  (IoT) applications and
machine-type communication (MTC) in 5G, where many   low-cost
sensors and devices transmit their information  with rough synchronization (or possibly asynchronously) to data collecting
centers through 5G networks. Thus, a new waveform that has low
latency and low OOB emission and  is robust against
CFO and timing offset (TO) is
necessary. Among multi-carrier (MC)-based waveforms beyond OFDM,
there are Filter-Bank MC (FBMC), Universal Filtered MC (UFMC),
GFDM and so on as candidates for the 5G new waveform to meet such
requirements\cite{Wunder&etal:14ComMag}.  FBMC can have low OOB
emission and robustness against TO and CFO by designing a long
prototype filter extending over multiple symbol intervals based on
Heisenberg's uncertainty principle
\cite{Chang:66Bell,Saltzberg:67COM,Flochetal:95PIEEE,KozekMolish:98JSAC,StrohmerBeaver:03COM},
but it has long delay and should have receiver equalization
instead of simply adding cyclic prefix (CP) to cope with channel's
delay spread. On the other hand, UFMC
\cite{Vakilianetal:13Globecom,Mukherjeeetal15IWCMC} and GFDM
\cite{Michailow&etal:14COM,Fettweis&etal:09VTC,Michailow&etal:12ISWCS,Gaspar&etal:13VTC}
respectively perform subband filtering and circular convolution
filtering  within one symbol to achieve such goals. Thus, they
have the advantage of low delay, and equalization can be avoided
by simply attaching  a CP to each symbol since processing is
contained within one symbol.  In this paper, we consider GFDM
 which is an
extension of single-carrier FDM (SC-FDM)   used  for 4G uplink and has an additional advantage
of low peak-to-average-power ratio (PAPR) due to its subsymbol structure, and  provide a
systematic framework for optimal filter design for GFDM based on a
rigorous rate analysis for GFDM under several receivers including
linear matched-filter (MF), ZF and MMSE receivers.

{\em Summary of Results:}  ~ The results of this paper are
summarized as follows:

$\bullet$ A rigorous rate analysis on GFDM systems  is provided and the transmission rate is derived as an
explicit function of GFDM filter coefficients in the AWGN channel
case.  It is shown that in  AWGN channels with no CFO, Dirichlet filters are rate-optimal under linear ZF
or MMSE receivers.

$\bullet$ The OOB radiation minimization
problem under a rate loss constraint is formulated, and it is shown that significant OOB emission reduction can be achieved without rate loss  by designing the GFDM filter properly if nonlinear successive interference cancellation (SIC) is assumed at the receiver.

$\bullet$  A framework for GFDM filter design for rate maximization in uplink GFDM
networks perturbed by CFO is presented and it is shown that proper
GFDM filtering can yield better rate performance than Dirichlet filters in this case.

$\bullet$  Finally, joint design frameworks of GFDM filter and
window  are  proposed to enhance the performance of GFDM further.


 {\it Notations:} ~  Vectors and matrices are written in boldface with matrices in capitals. All vectors are column vectors. For a matrix $\Abf$, $\Abf^T$, $\Abf^*$ $\Abf^H$, $[\Abf]_{i,j}$, and $\Abf(:,k)$ indicate the transpose,  complex conjugate,  conjugate  transpose,  $(i,j)$-th element, and  $k$-th column of $\Abf$, respectively.
$\Ibf_n$ stands for the identity matrix of size $n$ and
${\mathbf{1}}_n$ stands for the vector of size $n$ whose entries
are all one. $\mathrm{diag}(\Abf_1,\cdots, \Abf_n)$ is the block
diagonal matrix with diagonal elements $\Abf_1,\cdots, \Abf_n$.
$\Wbf_M$ is the $M$-point DFT matrix such that
$[\Wbf_M]_{i,j}=\frac{1}{\sqrt{n}}e^{-\iota 2\pi (i-1)(j-1)
/M},~i,j=1,2,\cdots,n$ with $\iota:=\sqrt{-1}$. $\delta_{m,n}$ is
the Kronecker delta, i.e., $\delta_{m,n}=1$ if $m=n$ and
$\delta_{m,n}=$ if $m\ne n$. The notation $\xbf\sim
\mathcal{CN}(\mubf,\Sigmabf)$ means that $\xbf$ is
circularly-symmetric complex Gaussian distributed with mean vector
$\mubf$ and covariance matrix $\Sigmabf$. $\Ebb\{\cdot\}$ denotes
the expectation.

\section{System Model}
 \label{sec:systemmodel}

 We first consider  GFDM transmission without windowing by postponing  joint design of GFDM filter and window  to Section \ref{sec:jointDesign}.
  Let
 $\sbf[i]=[s_0[i],\cdots,s_{N-1}[i]]^T$ be the data
 vector composed of the $N$ complex data subsymbols $s_0[i],\cdots,s_{N-1}[i]$ carried by the $i$-th GFDM symbol. Since
each symbol processing is independent and identical in GFDM, we
 consider one symbol time interval with  the symbol time index omitted. In GFDM, the data vector $\sbf$ is
decomposed into $K$ subvectors each with size $M$ as
\begin{equation}
\sbf=[s_{0},\cdots,s_{N-1}]^T=[\sbf_0^T,\cdots,\sbf_{K-1}^T]^T,
\end{equation}
 where
  $K$ is the number of  subcarriers,  $M$ is the
number of subsymbols carried by one subcarrier such that $N=MK$, and
$\sbf_k=[s_{kM},\cdots,s_{kM+M-1}]^T$  for $k=0,\cdots,K-1$. In
GFDM, subvector $\sbf_k$ is carried by the $k$-th subcarrier based
on time-domain filtering. That is, $s_{kM+m}$ is carried on the
$m$-th subsymbol in the $k$-th subcarrier with the shaping pulse  \cite{Michailow&etal:14COM}%
\begin{equation}
\tilde{g}_{k,m}[n]=\tilde{g}[n-mK] e^{\iota 2\pi k
n/K},~n=0,\cdots,N-1,
\end{equation}
where  $n$ is the time sample  or chip index, and
$\tilde{g}[n-mK]$ is the $mK$-sample {\em circularly shifted} version of
the time-domain GFDM filter $g[n]$. Thus, the
transmit signal of GFDM for one symbol time interval is expressed  as \cite{Michailow&etal:12ISWCS,Michailow&etal:14COM}%
\begin{equation} \label{eq:xknfrontModelpart}
x[n]=\sum_{k=0}^{K-1}x_k[n], ~~~n=0,1,\cdots,N-1,
\end{equation}
where $x_k[n]=\sum_{m=0}^{M-1}\tilde{g}_{k,m}[n]s_{kM+m}=\sum_{m=0}^{M-1}\tilde{g}[n-mK] e^{\iota 2\pi kn/K} s_{kM+m}$.

\subsection{The Matrix Transmit Signal Model}
\label{subsec:matrix model}

The transmit signal  \eqref{eq:xknfrontModelpart} can be written in a matrix
form as
\begin{equation}\label{eq:xbf}
\xbf= \Phibf\sbf,
\end{equation}
where $\xbf=[x[0],\cdots,x[N-1]]^T$ and  $\Phibf$ is an $N\times
N$ matrix given by $[\Phibf]_{n,kM+m}=\tilde{g}_{k,m}[n]$.  The
matrix $\Phibf$ can be expressed as the product of several
relevant matrices \cite{Michailow&etal:12ISWCS}. First, note that
 $x_k[n]$ in \eqref{eq:xknfrontModelpart} can be rewritten as
\begin{align}
x_k[n]&=\left[\left(\sum_{m=0}^{M-1} \delta[n-mK]\right)\circledast
g[n]\right] e^{\iota 2\pi kn /K}s_{kM+m}, \label{eq:subsig}
\end{align}
where $\circledast$ denotes the $N$-point circular convolution.
Furthermore, the signal \eqref{eq:subsig} can be rewritten based
on the properties of discrete Fourier transform (DFT) as
\eqref{eq:modelconv} \cite{Michailow&etal:12ISWCS}.
\begin{figure*}
\begin{align}
x_k[n]&=\textrm{IDFT}_{N}\left(\textrm{DFT}_{N}\left([(\sum_{m=0}^{M-1}s_{kM+m}\delta[n-mK])\circledast g[n]]e^{\iota 2\pi kn/K}\right)\right) \nonumber\\
&=\textrm{IDFT}_{N}\left(\textrm{DFT}_{N}\left(\sum_{m=0}^{M-1}s_{kM+m}\delta[n-mK]\right)\cdot
\textrm{DFT}_{N}(g[n])\circledast\textrm{DFT}_{N}\left(e^{\iota 2\pi kn/K}\right)\right). \label{eq:modelconv}
\end{align}
\end{figure*}
Note in  \eqref{eq:modelconv} that
\[
\textrm{DFT}_{N}\left(\sum_{m=0}^{M-1}s_k[m]\delta[n-mK]\right)=\frac{1}{\sqrt{K}}\left.\left[\begin{array}{c}\mathbf{W}_M\mathbf{s}_k\\\vdots\\
\mathbf{W}_M\mathbf{s}_k\end{array}\right]\right\}N,
\]
 $\textrm{DFT}_{N}(g[n])$ is the $N$-point frequency response
of the GFDM filter $g[n]$,  and convolution with
$\textrm{DFT}_{N}\left(e^{\iota 2\pi kn/K}\right)$ corresponds to
$kM$-sample circular shift with scaling $\sqrt{N}$ in the
frequency domain. Applying the above observations to
\eqref{eq:modelconv}, we can rewrite
 the GFDM transmit signal  \eqref{eq:xbf}  as
$\xbf=\Wbf_N^H\sum_{k=0}^{K-1}\tilde{\Pbf}_k \tilde{\Gammabf} \tilde{\Rbf} \Wbf_M \sbf_k$,
where $\tilde{\Rbf}$ is the repetition matrix given by
$\tilde{\Rbf}={\underbrace{[\Ibf_{M},\cdots,\Ibf_M]}_{K}}^T$, $\tilde{\Gammabf}$
is the $N\times N$ diagonal frequency-domain filtering matrix
containing $\sqrt{M}\textrm{DFT}_{N}(g[n])$ as its diagonal elements, and $\tilde{\Pbf}_k$ is the $N \times N$ permutation matrix implementing $kM$-sample circular shifting\cite{Michailow&etal:12ISWCS}.

In typical GFDM systems, the frequency response of the pulse
shaping filter $g[n]$ is designed to be  zero except $LM$ samples.
In this case,  $\tilde{\Rbf}$, $\tilde{\Gammabf}$ and $\tilde{\Pbf}$ in the expression $\xbf=\Wbf_N^H\sum_{k=0}^{K-1}\tilde{\Pbf}_k \tilde{\Gammabf} \tilde{\Rbf} \Wbf_M \sbf_k$ can be replaced with the
reduced-size $L$ times repetition matrix $\Rbf$ , the $LM\times
LM$ diagonal frequency-domain filtering matrix $\Gammabf$, and the $N\times LM$ permutation matrix
$\Pbf_k$, respectively.  In the example of a raised cosine (RC) or square-root RC (RRC) filter
$g[n]$  with a roll-off factor $\alpha \in [0,1]$, the repetition
factor $L$ is two and signal mixing for desired properties by GFDM
occurs only with two adjacent subcarriers for each subcarrier
\cite{Michailow&etal:14COM}. In this paper, we consider the
mostly-considered case of $L=2$  as in
\cite{Michailow&etal:12ISWCS,Michailow&etal:14COM} from here on. In this case,
we have \cite{Michailow&etal:12ISWCS}
\begin{equation} \label{eq:low}
\xbf=\Wbf_N^H\sum_{k=0}^{K-1}\Pbf_k \Gammabf \Rbf \Wbf_M \sbf_k,
\end{equation}
where
$\Rbf=[\Ibf_M ~\Ibf_M]^T$,
$
\mathbf{\Gamma}=\left[\begin{array}{cc}
\Gammabf^{(f)}&\mathbf{0}\\
\mathbf{0}&\Gammabf^{(r)}\end{array}\right]
=\mbox{diag}(\gamma_0,\gamma_1,\cdots,\gamma_{2M-1})$,
 and $\Pbf_k$ is the $N\times 2M$ subcarrier mapping matrix for the $k$-th subcarrier which is
 given by
\begin{equation}
\Pbf_k=\left[
\begin{array}{cccccc}
\mathbf{0}_{M}&\cdots&\mathbf{0}_{M}&\underbrace{\Ibf_{M}}_{k-\mbox{th}}&\cdots&\mathbf{0}_{M}\\
\mathbf{0}_{M}&\cdots&\Ibf_{M}&\mathbf{0}_{M}&\cdots&\mathbf{0}_{M}
\end{array}\right]^T.
\end{equation}
Here,  $\Gammabf^{(f)}$ and $\Gammabf^{(r)}$ are diagonal matrices with
size $M$, and
$\gamma_0,\cdots, \gamma_{2M-1}$ are the {\em design variables} in
GFDM filter design. The data model \eqref{eq:low} can further be
expressed as
\begin{equation} \label{eq:TXsigBigMatrix}
\xbf = \Wbf_N^H   \Pbf \mbox{diag}(\Fbf,\cdots,\Fbf)\sbf,
\end{equation}
where
$\Pbf =[\Pbf_0, \cdots, \Pbf_{K-1}]$  and $\Fbf = \Gammabf\Rbf\Wbf_M$.
 Thus, $\Phibf$ in \eqref{eq:xbf} is given by
\begin{equation}   \label{eq:PsibfinModel}
\Phibf = \Wbf_N^H   \Pbf \mbox{diag}(\Fbf,\cdots,\Fbf).
\end{equation}

 Note that GFDM with $L=1$ subsumes  OFDM, SC, and SC-FDM.
When $L=1$, $K=N$, and $M=1$,  the corresponding GFDM is
OFDM. When $L=1$, $K=1$, and $M=N$,   the corresponding GFDM is  SC. When  $L=1$,
$1< M < N$, $\Gammabf=\Ibf_M$ (i.e., the filter $g[n]$ is a Dirichlet
filter which is the RC filter with roll-off factor $\alpha=0$),
and $\Pbf_k=[\mathbf{0},\cdots,\Ibf_M,\cdots,\mathbf{0}]^T$,
the corresponding GFDM reduces to SC-FDM\cite{Michailow&etal:14COM}.

We assume that the data subvectors $\sbf_k$, $k=0,\cdots,K-1$ are
zero-mean independent Gaussian random vectors  with $\sbf_k \sim
\Cc\Nc({\mathbf{0}}, P_s \Ibf)$, where $P_s$ is the data subsymbol
power.  Then, the total transmit power of one GFDM symbol is
obtained from \eqref{eq:low} as
\begin{align}
P_t &= \mbox{tr}(\xbf\xbf^H)  = P_s\sum_{k=0}^{K-1} \mbox{tr} ( \Pbf_k \Gammabf \Rbf \Wbf_M \Wbf_M^H \Rbf^H \Gammabf^H \Pbf_k^H)  \nonumber\\
&= K  P_s \sum_{m=1}^{2M-1} |\gamma_m|^2,
\end{align}
where the fact that $\Wbf_M \Wbf_M^H = \Ibf$ and
$\Pbf_k^H\Pbf_k=\Ibf$ and the structure of $\Rbf$ are used. We set
$P_t=N P_s$ to have a normalized filter power constraint
\begin{equation} \label{eq:filterPowerConst}
\sum_{m=1}^{2M-1} |\gamma_m|^2=M.
\end{equation}

\subsection{The Channel  and  Receiver Model}
\label{subsec:rec}

We assume that a CP of size $N_{\mathrm{cp}}$ samples is added to
the transmit signal vector $\xbf$ in \eqref{eq:TXsigBigMatrix} in
the time domain to yield  the CP-added transmit signal
$\bar{\xbf}=[x[N-N_{\mathrm{cp}}],\cdots, x[N-1],x[0],\cdots,x[N-1]]^T$.
This CP-added signal vector  $\bar{\xbf}$ is transmitted through a
multipath fading channel with an $N_{\mathrm{cp}}$-tap finite
impulse response (FIR) $\hbf=[h[0],\cdots,h[N_{\mathrm{cp}}-1]]^T$
 and is received at the receiver with
AWGN to yield the received
signal vector
$\bar{\ybf}=[y[-N_{\mathrm{cp}}],\cdots,y[-1],y[0],\cdots,y[N-1]]^T$.
At the receiver, the first $N_{\mathrm{cp}}$ samples of
$\bar{\ybf}$ corrupted by  inter-block interference
are removed. Thus, the CP-portion-removed received signal vector
$\ybf=[y[0],\cdots,y[N-1]]^T$ is given by
\begin{equation}\label{eq:ybfinRXmodel}
\ybf=\Hbf\xbf+\nbf=\Hbf\Phibf\sbf+\nbf,
\end{equation}
where $\Hbf$ is the $N\times N$ circulant channel matrix with the
first column given by $[\hbf^T ~\mathbf{0}^T]^T$, and the noise
vector $\nbf \sim \mathcal{CN} (\mathbf{0},\sigma_n^2 \Ibf)$ with
the noise variance $\sigma_n^2$.\footnote{Note that the assumption
that the whole vector $\xbf$, which is the sum of all subcarrier
signals, goes through the single channel $\hbf$ is relevant to
point-to-point communication scenarios. An uplink network scenario
will be considered later in Section \ref{subsec:CFOrobustness}.}

We consider standard receivers.  In  linear receivers, an estimate of the symbol vector for further processing is
obtained as
$\hat{\sbf}= \Lbf \ybf$.
The MF receiver $\Lbf = \Phibf^H$
 is simple and aims at maximizing the  signal-to-noise
ratio (SNR) for each subsymbol without considering  interference. In
the case of OFDM and SC-FDM, the MF receiver $\Lbf=\Phibf^H$
diagonalizes the circulant channel matrix $\Hbf$, since
$\Wbf_N\Hbf\Wbf_N^H$ is a diagonal matrix, and separates the
subcarrier channels without inter-subcarrier interference.  In nontrivial
GFDM, however, symbol mixing is made intentionally to achieve
certain design goals as seen in \eqref{eq:low} with the repetition
matrix $\Rbf$ and the summation operation. Thus, the simple MF
receiver experiences inter-subcarrier interference  and this
inter-subcarrier interference  limits the performance of the MF
receiver in GFDM. To eliminate the inter-subcarrier interference, SIC can be employed on top of  the MF receiver for GFDM, or   the linear ZF
receiver $\Lbf=(\Hbf\Phibf)^{-1}$ or
 the linear MMSE receiver
 $\Lbf=((\sigma_n^2/P_s)
\Ibf_N + \Phibf^H\Hbf^H\Hbf\Phibf)^{-1} \Phibf^H\Hbf^H$ can be used with reduced complexity.

\section{Rate Analysis for GFDM in the AWGN  Case}
\label{sec:filterdesignrate}

In this section, we analyze the rate of GFDM with the receivers
considered in Section \ref{subsec:rec} and derive a rate-optimal GFDM
filter for linear ZF or MMSE receivers in the  AWGN channel case
with no CFO, which is one of the most important channel models in
communication.

\vspace{-1em}

\subsection{The MF  or MF/SIC Receiver Case}  \label{subsec:MFreceiver}

The  signal estimate of the MF receiver  is given by
\begin{equation} \label{eq:MForg}
\hat{\sbf}=\Phibf^H\ybf = \Phibf^H \Hbf \Phibf
\sbf + \Phibf^H\nbf.
\end{equation}
From \eqref{eq:TXsigBigMatrix},  \eqref{eq:PsibfinModel} and
\eqref{eq:ybfinRXmodel}, the estimated signal vector at the $k$-th
subcarrier can be written as
\begin{equation}\label{eq:MF}
\hat{\sbf}_k= \Fbf^H\Pbf_k^H\Lambdabf_H\Pbf \mathrm{diag}(\Fbf,\cdots,\Fbf)\sbf + \Fbf^H\Pbf_k^H\Wbf_N\nbf,
\end{equation}
where $\Lambdabf_H=\Wbf_N\Hbf\Wbf_N^H$ is the $N\times N$ diagonal
matrix whose diagonal elements are the eigenvalues of the
circulant channel matrix $\Hbf$.

In the AWGN channel case, we have  $\Hbf=\Ibf$ and $\ybf = \Phibf
\sbf + \nbf$ from \eqref{eq:ybfinRXmodel}, and the sum capacity of
the data model $\ybf = \Phibf \sbf + \nbf$ is   given by
\cite{Telatar99}
\begin{equation} \label{eq:Rmax}
\sum_{n=0}^{N-1} \log \left(1+ \frac{P_s}{\sigma_n^2} \xi_n^2\right) \le N \log \left(1+ \frac{P_s}{\sigma_n^2}\right)=:R_{max},
\end{equation}
 where  $\xi_n$ is the $n$-th singular value of the matrix $\Phibf$ and the inequality  in \eqref{eq:Rmax} is obtained by applying Jensen's inequality to the concave function $f(x)=\log(1+x)$ with the power  constraint  $P_t=\mbox{tr}(\xbf\xbf^H)=P_s\mbox{tr}(\Phibf\Phibf^H)= P_s||\Phibf||_F^2 = P_s \sum_{n=0}^{N-1} \xi_n^2 \le  NP_s$, i.e., $\frac{1}{N}\sum_{n=0}^{N-1} \xi_n^2 \le 1$.  The upper bound
 in \eqref{eq:Rmax} is achieved  when $\xi_0=\cdots=\xi_{N-1}=1$, equivalently, $\Phibf^H\Phibf = \Phibf\Phibf^H  = \Ibf$, i.e., the
full orthonormality  among the pulses $\tilde{g}_{k,m}[n]=[\Phibf]_{n,kM+m}$ carrying $s_{kM+m}$, $m=0,1,\cdots,M-1$,
$k=0,1,\cdots,K-1$ is satisfied.  (A similar observation was made previously in  \cite{Ngo&Larsson&Marzetta:13COM} for a MIMO context.)

 Indeed,
 if the
full orthonormality among the pulses $\tilde{g}_{k,m}[n]$ carrying $s_{kM+m}$, $m=0,1,\cdots,M-1$,
$k=0,1,\cdots,K-1$ were preserved, then the MF symbol SNR would be
given from \eqref{eq:MForg} with $\Hbf=\Ibf$ by
$\mbox{SNR}_{k,m} =\frac{P_s}{\sigma_n^2}\sum_n
|\tilde{g}_{k,m}[n]|^2 =\frac{P_s}{\sigma_n^2}\sum_n
|g[n]|^2 \stackrel{(a)}{=}\frac{P_s}{\sigma_n^2}\sum_{l=0}^{M-1}\frac{1}{M}
(|\gamma_l|^2+|\gamma_{M+l}|^2)\stackrel{(b)}{=}\frac{P_s}{\sigma_n^2}$,
which is the {\em ideal MF SNR bound}, where step (a) is due to
Parseval's theorem and step (b) is due to the filter power
constraint \eqref{eq:filterPowerConst}, and the corresponding sum rate would be given by $R_{max}$ in  \eqref{eq:Rmax}.
However, when a non-trivial GFDM filter $\Gammabf$  with $L= 2$ is
applied, intentional inter-subcarrier mixing  occurs, the orthogonality among $\tilde{g}_{k,m}[n]$ is broken, and
inter-subcarrier interference exists. In this case,  the
signal-to-interference-plus-noise ratio (SINR) for symbol
$s_{kM+m}$  degrades from the ideal MF SNR bound and
is given in the following theorem.

\begin{theorem}  \label{lem:MFGFDMrate}
Under the  AWGN channel $\Hbf=\Lambdabf_H=\Ibf_N$ with the
transmit power $P_t=NP_s$, the
SINR of the $m$-th subsymbol
of the $k$-th subcarrier in GFDM with the MF receiver is given by
$\mathrm{SINR}_{k,m}^{MF}(\gammabf)=\frac{P_s}{a(\gammabf)P_s+\sigma_n^2}$,
where $\gammabf=[\gamma_0,\cdots,\gamma_{2M-1}]$, and $a(\gammabf)=\frac{1}{M^2}\sum_{p\neq m,p=0}^{M-1}\left|\sum_{l=0}^{M-1}e^{\iota 2\pi (m-p)l/M} (|\gamma_l|^2+|\gamma_{M+l}|^2)\right|^2+\frac{2}{M^2}\sum_{p=0}^{M-1}\left|\sum_{l=0}^{M-1}e^{\iota 2\pi (m-p)l/M}\gamma_l\gamma_{M+l}^*\right|^2$.
\end{theorem}

{\em Proof:} See Appendix.

 Then, the sum rate of GFDM with the MF receiver is given by
$R_{MF}(\gammabf)=\sum_{k=0}^{K-1}\sum_{m=0}^{M-1} \log
(1+\mathrm{SINR}_{k,m}^{MF})$,
where  $\mathrm{SINR}_{m,k}^{MF}$ is given in Theorem \ref{lem:MFGFDMrate}
and depends on the GFDM filter $\gammabf$. With a non-trivial GFDM
filter $\gammabf$, e.g., the  RRC filter with a non-zero roll-off
factor\cite{Michailow&etal:14COM}, the rate performance of the MF
receiver for GFDM  saturates at high SNR due to the intentional
inter-subcarrier interference, as shown in Fig.
\ref{fig:ratelossComp}. In order to eliminate the inter-subcarrier
interference for the MF receiver, SIC can be applied on top of the
MF  \cite{Gaspar&etal:13VTC,Michailow&etal:14COM}.
 Note from Proof of Theorem
\ref{lem:MFGFDMrate} that the term $a(\gammabf)P_s$ in the denominator of
the right-hand side (RHS) of the $\mathrm{SINR}_{k,m}^{MF}(\gammabf)$ expression in Theorem \ref{lem:MFGFDMrate}
  is the interference
from other data subsymbols.  Hence, if   SIC is performed until
the interference is  removed, then  the interference term $a(\gammabf)P_s$
disappears and the best subsymbol SNR in this case is given by
${P_s}/{\sigma_n^2}$, yielding  the maximum rate \eqref{eq:Rmax}.
(For implementation of SIC in GFDM, see
\cite{Gaspar&etal:13VTC}.)

Note that in the AWGN case of $\Hbf=\Ibf$, the MF output is given
by $\hat{\sbf}=\Phibf^H\ybf = \Phibf^H\Phibf \sbf + \Phibf^H\nbf$
from \eqref{eq:MForg} and the corresponding subsymbol SINR is
given by $\mathrm{SINR}_{k,m}^{MF}(\gammabf)=\frac{P_s}{a(\gammabf)P_s+\sigma_n^2}$. Hence, the orthonormality condition
$\Phibf^H\Phibf=\Ibf$ to achieve  the upper bound in
\eqref{eq:Rmax} is equivalent to $a(\gammabf)=0$. The explicit
necessary and sufficient condition for $a(\gammabf)=0$ is not
straightforward to obtain, but a sufficient condition is given by
the Dirichlet filter $\gamma_0=\cdots=\gamma_{M-1}=1$ and
$\gamma_M=\cdots=\gamma_{2M-1}=0$. We shall discuss this condition
in more detail in the next subsection.

\subsection{Maximum Rate Filtering for the ZF or MMSE Receiver}
\label{sec:ZFfilterdesignrate}

 In this section, as an alternative to the MF/SIC receiver, we  consider the  ZF or MMSE receiver, which eliminates
inter-subcarrier and inter-subsymbol interference with simple
linear processing, and  derive an optimal GFDM
filter that maximizes the data rate for GFDM under the ZF or MMSE
receiver.

In the case of the ZF receiver, the estimated signal vector is given from \eqref{eq:PsibfinModel} and
\eqref{eq:ybfinRXmodel} by
\begin{align}
\left[\begin{array}{c}\hat{\sbf}_0\\\vdots\\\hat{\sbf}_{K-1}\end{array}\right]&=(\Hbf\Wbf_N^H\Pbf\mathrm{diag}(\Fbf,\cdots,\Fbf))^{-1}\ybf\nonumber\\
&=(\Wbf_N^H\Wbf_N\Hbf\Wbf_N^H\Pbf\mathrm{diag}(\Fbf,\cdots,\Fbf))^{-1}\ybf\nonumber\\
&=(\Pbf\mathrm{diag}(\Fbf,\cdots,\Fbf))^{-1}\Lambdabf_H^{-1}\Wbf_N\ybf \label{eq:ZFprocess}\\
&=\left[\begin{array}{c}\sbf_0\\\vdots\\\sbf_{K-1}\end{array}\right]+(\Pbf\mathrm{diag}(\Fbf,\cdots,\Fbf))^{-1}\Lambdabf_H^{-1}\tilde{\nbf},
\nonumber
\end{align}
where $\tilde{\nbf}=\Wbf_N\nbf\sim\mathcal{CN}
(\mathbf{0},\sigma_n^2\mathbf{I}_N)$.

\vspace{0.5em}
\begin{remark}
Note from \eqref{eq:ZFprocess} that the ZF receiver processing can
be done efficiently with low complexity.
  We first apply the $N$-point DFT to $\ybf$,
multiplication of diagonal $\Lambdabf_H^{-1}$, which is simple
elementwise scaling, and multiplication of
$(\Pbf\mathrm{diag}(\Fbf,\cdots,\Fbf))^{-1}$. The last step can
also be performed with low complexity by reformulating
$(\Pbf\mathrm{diag}(\Fbf,\cdots,\Fbf))^{-1}$ with block DFT
matrices. See Proof of Theorem \ref{lem:ZFGFDMrate} and  Remark
\ref{remark:appendLowComplex} in Appendix for detail.
\end{remark}
\vspace{0.5em}

In the ZF receiver case, the signal part is completely separated
but the noise is enhanced. The noise enhancement reduces the
effective channel gain of each separate parallel Gaussian channel
provided by ZF processing \cite{JiangVaranasiJianLi:11IT}. The rate of
GFDM with the ZF receiver is the sum of the rates of the $N$ parallel
Gaussian channels and is given by the following theorem.

\vspace{0.5em}

\begin{theorem} \label{lem:ZFGFDMrate} In the general FIR channel case,  the sum rate of GFDM with the ZF receiver is given by {\small
\begin{equation} \label{eq:ZFcap}
R_{ZF}(\gammabf)=M\sum_{k=0}^{K-1}\log\left(1+\dfrac{P_s/\sigma_{n}^2}{\sum_{p=0}^{K-1}
\sum_{q=0}^{M-1}\left|\sum_{l=0}^{K-1}
\dfrac{c_{k,p,q,l}}{\gamma_{q}+d_l\gamma_{M+q}} \right|^2}\right),
\end{equation}}
where $c_{k,p,q,l}=\dfrac{1}{\sqrt{M}}\dfrac{1}{K}e^{\iota 2\pi
(k-p )l/K}[\Lambdabf_{H}^{-1}]_{pM+q,pM+q}$ and $d_{l}=e^{\iota
2\pi l /K }$.  In the AWGN channel case,  the sum rate of GFDM with the ZF receiver is given by {\small
\begin{equation}\label{eq:ZFawgnrate}
R_{ZF}(\gammabf)=N\log\left(1+\dfrac{P_s/\sigma_n^2}{\dfrac{1}{N}\sum_{l=0}^{K-1} \sum_{q=0}^{M-1}\dfrac{1}{|\gamma_{q}+d_l\gamma_{M+q}|^2}}\right).
\end{equation}}
\end{theorem}

{\em Proof:} See Appendix.

\vspace{0.5em}

Using Theorem \ref{lem:ZFGFDMrate}, we can now formulate the problem
of finding an optimal GFDM filter to maximize the rate of GFDM
with the ZF receiver. In particular,  in the AWGN channel case, by
exploiting the monotonicity of the logarithm function and the
function $f(x)=1/x$ in  \eqref{eq:ZFawgnrate}, the optimization
problem reduces to the following problem:

\begin{problem} \label{prob:ZFrateMaximizeaa}
\begin{equation}
\begin{array}{cl}
\underset{\gammabf}{\arg \min}&\sum_{l=0}^{K-1} \sum_{q=0}^{M-1}\dfrac{1}{|\gamma_{q}+d_l\gamma_{M+q}|^2}\\
\mathrm{ s.t. }&\sum_{m=0}^{2M-1}|\gamma_m|^2=M.
\end{array}
\end{equation}
\end{problem}

 A solution to Problem  \ref{prob:ZFrateMaximizeaa} is
given by the following theorem.

\vspace{0.5em}

\begin{theorem} \label{theo:DiriOpt}
In the AWGN channel case,  the Dirichlet filter (i.e.,
$\gamma_0=\cdots=\gamma_{M-1}= 1$ and
$\gamma_{M}=\cdots=\gamma_{2M-1}=0$) maximizes the  rate of GFDM with the ZF receiver
and the corresponding rate is $N\log
\left(1+\frac{P_s}{\sigma_n^2}\right)$. Furthermore, any
filter  cannot have a larger rate  than the Dirichlet
filter for GFDM with the ZF receiver.
\end{theorem}

{\em Proof:} See Appendix.

\vspace{0.5em}

 Theorem \ref{theo:DiriOpt} states that for given $M$ and
$K(>1)$, SC-FDM is rate-optimal among GFDM when the ZF receiver is
used at the receiver, since GFDM with the Dirichlet filter is SC-FDM.

Now consider the MMSE receiver case.  The  MMSE receiver in the
AWGN channel case  is given by $\Lbf=((\sigma_n^2/P_s) \Ibf +
\Phibf^H\Phibf)^{-1} \Phibf^H$.  In the MMSE receiver case,
computing the mutual information $I(\sbf;\hat{\sbf}=\Lbf\ybf)$, we have
the sum rate  given by
\begin{equation} \label{eq:sumrateeffectiveMMSE}
R_{MMSE}(\gammabf) = \sum_{k=0}^{K-1}\sum_{m=0}^{M-1} \log ( 1
+\rho_{k,m,MMSE}(\gammabf)),
\end{equation}
where the effective output SNR for each subchannel is given by
\cite{JiangVaranasiJianLi:11IT}
\begin{equation}\label{eq:effectiveMMSE}
\rho_{k,m,MMSE}(\gammabf)=\dfrac{P_s/\sigma_n^2}{\left[\left(\Phibf^H\Phibf+\frac{\sigma_n^2}{P_s}\Ibf\right)^{-1}\right]_{kM+m,kM+m}}-1.
\end{equation}
Based on \eqref{eq:sumrateeffectiveMMSE} and
\eqref{eq:effectiveMMSE}, the sum rate of GFDM with the MMSE receiver is derived in
the following theorem:

\vspace{0.5em}

\begin{theorem}  \label{theo:MMSErate}
In the AWGN channel case, the sum rate of GFDM with the MMSE receiver
is given by
\begin{equation}\label{eq:MMSEawgnrate}
R_{MMSE}(\gammabf)=N\log\left(\dfrac{P_s/\sigma_n^2}{D(\gammabf)}\right),
\end{equation}
where $D(\gammabf)=\dfrac{1}{N}\sum_{l=0}^{K-1}
\sum_{q=0}^{M-1}\dfrac{1}{|\gamma_{q}+d_l\gamma_{M+q}|^2+\sigma_n^2/P_s}$.
\end{theorem}

{\em Proof:} See Appendix.

\vspace{0.5em}

In a similar way to the ZF case,  the problem of filter
optimization for GFDM with the MMSE receiver in the AWGN channel
case is formulated based on  Theorem  \ref{theo:MMSErate} as
follows:

\begin{problem} \label{prob:rateMaximizeMMSE}
\begin{equation}
\begin{array}{cl}
\underset{\gammabf}{\arg \min}&\sum_{l=0}^{K-1} \sum_{q=0}^{M-1}\dfrac{1}{|\gamma_{q}+d_l\gamma_{M+q}|^2+\sigma_n^2/P_s}\\\mathrm{ s.t. }&\sum_{m=0}^{2M-1}|\gamma_m|^2=M.
\end{array}
\end{equation}
\end{problem}

A solution to Problem  \ref{prob:rateMaximizeMMSE} is given by the following theorem.

\vspace{0.5em}
\begin{theorem} \label{theo:DiriOptMMSE}
In the AWGN channel case,  the Dirichlet filter is  rate-optimal for GFDM with the MMSE receiver and the corresponding rate is $N\log \left(1+\frac{P_s}{\sigma_n^2}\right)$. Furthermore, any
filter  cannot have a larger rate  than the Dirichlet
filter for GFDM with the MMSE receiver.
\end{theorem}

{\em Proof:} See Appendix.

\vspace{0.5em}

As seen in Section  \ref{subsec:MFreceiver}, with the nonlinear
MF/SIC receiver, any filter satisfying the power constraint is
rate-optimal for GFDM under the assumption of ideal SIC in the
AWGN channel case. However,  Theorems \ref{theo:DiriOpt} and
\ref{theo:DiriOptMMSE} state that with the linear ZF or MMSE
receiver, the Dirichlet filter, i.e., SC-FDM with no intentional
inter-subcarrier mixing, is rate-optimal for GFDM, and the
achievable rate is identical to the sum capacity upper bound,
$R_{max}$ in \eqref{eq:Rmax}. In the example of an RRC filter with
a non-zero roll-off factor\cite{Michailow&etal:14COM},  the sum
rate loss occurs in the AWGN channel case with a linear receiver,
as seen in Fig. \ref{fig:ratelossComp}.  Theorems
\ref{theo:DiriOpt} and \ref{theo:DiriOptMMSE}  follow our
intuition that when the channel itself does not cause any
inter-subcarrier interference due to the same inter-subcarrier
spacing as in  point-to-point communication situations,
intentionally induced inter-subcarrier interference by non-trivial
GFDM filtering will not be beneficial for SNR and hence for the
 rate. However, it will be shown later that even in the linear receiver case, SC-FDM is not optimal and non-trivial
GFDM filtering enabling a countermeasure against channel-made
inter-subcarrier interference is actually beneficial for the rate
when there exists channel-made inter-subcarrier interference due
to  CFO  as in uplink network situations.

\section{Filter Design Examples}
\label{sec:filterdesignExamples}

 In this section, we provide two design examples for GFDM. The
first design goal is to minimize the OOB radiation and the second
is to maximize the rate under CFO.

\subsection{Design for OOB Radiation Minimization}
\label{sec:filterdesignother}

OOB radiation minimization is  one of the major issues in MC
communication. Especially, in asynchronous bursty transmission of
MTC or IoT data based on subcarrier modulation, synchronization required for waveform orthogonality is
difficult to achieve and thus the spectral leakage of the waveform
itself should be low not to significantly interfere with the data at different
subcarriers\cite{Wunder&etal:14ComMag}.  For OOB radiation
minimization, we first need to express the PSD of the continuous-time transmit signal
 generated from the discrete-time signal $x[n]$ as a function of the filter
coefficients $\gamma_0,\cdots,\gamma_{2M-1}$. From
\eqref{eq:xknfrontModelpart}, the
discrete-time transmit signal $x[n]$ can be rewritten  as
\begin{equation}  \label{eq:OOBx[n]}
x[n]=\sum_{k=0}^{K-1}e^{\iota 2\pi kn/K}
\sum_{m=0}^{M-1}\tilde{g}[n-mK]s_{kM+m},
\end{equation}
where $\tilde{g}[n-mK]$ is given in terms of
$\gamma_0,\cdots,\gamma_{2M-1}$ as {\small
\begin{equation} \label{eq:tildegnmkggOOB}
\tilde{g}[n-mK]=\frac{1}{\sqrt{NM}}\sum_{q=0}^{M-1}e^{\iota
2\pi(n-Km)q/N}(\gamma_q+e^{-\iota 2\pi n/K}\gamma_{M+q}).
\end{equation}
} (See Appendix  for detail.)  Let $T_s$ be one subsymbol time
interval corresponding to $K$ time-domain samples and $T_b=MT_s$
be one GFDM symbol time interval.  Under the assumption that a
sample-level sinc interpolation filter is used, the discrete-time
signal
 \eqref{eq:OOBx[n]} is converted by changing $n/K$ to $t/T_s$ to  the continuous-time  signal
for one GFDM symbol interval $T_b$  as
\begin{equation}  \label{eq:CTsignal}
x_{T_b}(t)=\sum_{k=0}^{K-1}
e^{\iota2\pi kt/T_s}\sum_{m=0}^{M-1} {g_m(t)}s_{kM+m}, %
\end{equation}
where $ g_m(t)=\frac{1}{\sqrt{NM}} \sum_{q=0}^{M-1}e^{-\iota 2\pi
mq/M} e^{ \iota 2\pi qt/(MT_s)}(\gamma_q+e^{-\iota 2\pi
t/T_s}\gamma_{M+q})$ for $0\leq t \leq T_b$ and $g_m(t)=0$ for $ t
\notin [0, T_b]$. If the CP part is ignored for
simplicity\footnote{When the CP part is considered, a similar
derivation is obtained with an increased time interval including
the CP portion for the same $g_m(t)$. Note that $g_m(t)$ and
$x_{T_b}(t)$ in \eqref{eq:CTsignal} are periodic with period
$MT_s$. See Section \ref{sec:jointDesign}.}, then $x_{T_b}(t)$ is
repeated for each GFDM symbol interval $i$ with independent data
symbols $s_{kM+m}[i]$ for each GFDM symbol interval $i$, and this
continuous-time signal is a cyclo-stationary random process
\cite{Proakis:book}. Note in \eqref{eq:CTsignal} that the overall
signal $x_{T_b}(t)$ is the shifted sum of  the baseband signals
$\sum_{m=0}^{M-1} {g_m(t)}s_{kM+m}$, $k=0,1,\cdots,K-1$ since
 multiplication by $e^{\iota2\pi kt/T_s}$ is shift in the frequency domain by $k/T_s$ Hz.
Note also that  the PSD of the baseband signal $\sum_{m=0}^{M-1}
{g_m(t)}s_{kM+m}$ is the same for all $k=0,1,\cdots,K-1$, and is
given by \cite[p.573]{Proakis:book}
\begin{equation} \label{eq:PBBf}
P_{BB}(f) = \frac{P_s}{T_b} \sum_{m=0}^{M-1} |G_m(f)|^2,
\end{equation}
where $G_m(f)$ is the Fourier transform of $g_m(t)$ given by
$G_m(f)=\int_{-\infty}^{\infty} g_m(t) e^{-\iota 2\pi
ft}dt=\frac{1}{\sqrt{NM}}\sum_{q=0}^{M-1}e^{-\iota 2\pi
mq/M}(e^{-\iota 2\pi T_b f}-1) (\dfrac{\gamma_q}{\iota 2\pi
(q/T_b-f)} +\dfrac{\gamma_{M+q}}{\iota 2\pi (q/T_b-f -1/T_s)}).$
Then, the overall PSD  is given by
\begin{equation}  \label{eq:PBBfall}
P(f)=\sum_{k=0}^{K-1}
P_{BB}\left(f-\frac{k}{T_s}\right).
\end{equation}
In the considered case of the repetition factor $L=2$,  the
nominal frequency  range of $P_{BB}(f)$ is
$f\in[-\frac{1}{T_s},\frac{1}{T_s}]$ and thus the nominal
frequency range of the overall PSD $P(f)$ is
$f\in[-\frac{1}{T_s},\frac{K}{T_s}]$. Based on the results
in Section \ref{sec:filterdesignrate} we now formulate optimal filter
design problems for OOB radiation minimization as follows:

\begin{problem}[OOB radiation minimization for GFDM-MF/SIC]
\label{prob:oobmfsic}
\begin{equation}
\begin{array}{cl}
&\underset{\gammabf}{\arg \min}\underset{f\in[\frac{1}{T_s},\infty)}{\max}  P(-\frac{1}{T_s}-f)+P(\frac{K}{T_s}+f) \\
&~\mathrm{s.t. }~~ \sum_{m=0}^{2M-1}|\gamma_m|^2=M.
\end{array}
\end{equation}
\end{problem}
Here, in the MF/SIC receiver case,  no rate constraint is applied
since any filter satisfying the power constraint does not cause a
rate loss if ideal SIC on top of the MF receiver is assumed.
However, in the linear receiver case a rate constraint should be
included in the optimization since the OOB-optimal filter is not
the rate-optimal filter in general.

\begin{problem}[OOB radiation minimization for GFDM-ZF] \label{prob:oobZF}
\begin{equation}  \label{eq:OOBminZF}
\begin{array}{cl}
&\underset{\gammabf}{\arg \min}\underset{f\in[\frac{1}{T_s},\infty)}{\max} P(-\frac{1}{T_s}-f)+P(\frac{K}{T_s}+f)\\
&~\mathrm{s.t. }~~
\sum_{m=0}^{2M-1}|\gamma_m|^2=M \\
&~~~~~~~ R_{ZF}(\gammabf) \ge (1-\eta) N\log(1+P_s/\sigma_n^2),
\end{array}
\end{equation}
where $\eta\in[0,1]$ is the allowed rate loss factor compared to
the maximum possible rate and $R_{ZF}(\gammabf)$ is provided in
Theorem \ref{lem:ZFGFDMrate}.
\end{problem}

In Problems \ref{prob:oobmfsic} and \ref{prob:oobZF}, $1/T_s$,
which is the spacing between two adjacent subcarriers, is selected
for the transition bandwidth from the passband to the stopband,
but a different value can be used. In the case of the MMSE receiver,
the same problem can be formulated as Problem \ref{prob:oobZF}
with $R_{ZF}(\gammabf)$ replaced by $R_{MMSE}(\gammabf)$ given in
Theorem \ref{theo:MMSErate}. Numerical results on the design by Problems \ref{prob:oobmfsic} and \ref{prob:oobZF} are provided in Section \ref{sec:numOOB}.

\subsection{Design for Rate Maximization in General Channels with CFO}
\label{subsec:CFOrobustness}

In section \ref{sec:ZFfilterdesignrate}, it is shown that the
Dirichlet filter (i.e., SC-FDM) is rate-optimal for GFDM with the
linear ZF or MMSE receiver in the AWGN channel case with no CFO.
However, this is not valid in general channels especially when the
channel is perturbed by CFO. In this subsection, we provide the
second design example of maximizing the transmission rate in an
uplink scenario with CFO.

In the uplink GFDM case, we need to change the system model
considered in  Section \ref{sec:systemmodel}. In the uplink
GFDM case, we have simultaneously transmitting $K$ users with
independent local oscillators (LOs) and a common receiver with a
common LO in the network, and  each user is allocated to one
subcarrier carrying $M$ subsymbols. The transmit data of user $k$
is $\sbf_k$, and the transmit signal vector of user $k$ with size $N+N_{cp}$ samples is given by
\begin{equation}
\bar{\xbf}_k=\Omegabf_t\Wbf_N^H\Pbf_k \Gammabf\Rbf\Wbf_M \sbf_k,
\end{equation}
 where
$\Omegabf_t=\left[[\mathbf{0}~\Ibf_{N_{\mathrm{cp}}}]^T~\Ibf_{N}\right]^T$
is the CP addition matrix at the transmitter. The transmit
signal of user $k$ propagates through an FIR channel
$\hbf_k=[h_k[0],\cdots,h_k[N_{cp}-1]]^T$ to reach the common
receiver. In general, there exists CFO for each user $k$ because
of the discrepancy between the LO frequency of each user used in
upconversion and the LO frequency of the common receiver used in
downconversion.  Let $\epsilon_k=\frac{\hat{f}_k-f_k}{\Delta f}$
be the CFO of user $k$, where $\hat{f}_k$, $f_k$, and $\Delta f$
are the estimated carrier frequency of user $k$, the original
transmit carrier frequency of user $k$, and the frequency spacing
of two adjacent subcarriers, respectively.
The impact of CFO on the
baseband model is implemented simply by multiplying the diagonal  CFO
matrix
$
\bar{\Pibf}(\epsilon_k)=\mathrm{diag}(e^{-\iota2\pi\epsilon_k
N_{cp}},\cdots,e^{-\iota2\pi\epsilon_k},1,\cdots,e^{\iota2\pi\epsilon_k
(N-1)})$
 to the baseband channel output  $\hbf_k^T * \bar{\xbf}_k^T$  of user $k$\cite[eq.4]{LeeLouCioffi:04GC}.   Thus, the
received signal of GFDM with CFO after CP removal and DFT is expressed as
\begin{align}
\ybf_{CFO} &= \sum_{k=0}^{K-1}\Wbf_N \Pibf(\epsilon_k) \Omegabf_r \bar{\Hbf}_k
 \Omegabf_t\Wbf_N^H\Pbf_k
\Gammabf\Rbf\Wbf_M \sbf_k +\nbf, \nonumber\\
&=\sum_{k=0}^{K-1} \Dbf_k \Lambdabf_k\Pbf_k
\Gammabf\Rbf\Wbf_M \sbf_k +\nbf,\label{eq:CFOrec}
\end{align}
where $\bar{\Hbf}_k$ is
the $(N+N_{\mathrm{cp}})\times (N+N_{\mathrm{cp}})$ Toeplitz
channel filtering matrix for user $k$ with $[\hbf_k^T ~
\mathbf{0}^T]^T$ as the first column,  $\Omegabf_r=\left[\mathbf{0}~\Ibf_{N}\right]$ is  the CP removal
matrix,  $\Pibf(\epsilon_k)$ is the lower-right $N\times N$ submatrix of $\bar{\Pibf}(\epsilon_k)$, $\Dbf_k=\Wbf_N \Pibf(\epsilon_k)\Wbf_N^H$, and  $\Lambdabf_k=\Wbf_N\Omegabf_r \bar{\Hbf}_k \Omegabf_t\Wbf_N^H$. The $N\times N$ matrix $\Dbf_k$ with elements $[\Dbf_k]_{i,j}=\frac{\sin \pi (N\epsilon_k - (i-j))}{\sin (\epsilon_k - (i-j)/N)}e^{\pi(N-1)(\epsilon_k-(i-j)/N}$ of the noncentered Dirichlet kernel \cite{LeeLouCioffi:04GC} is not diagonal with $\epsilon_k \ne 0$   and induces inter-subcarrier interference.  The received signal $\ybf_{CFO}$ in \eqref{eq:CFOrec} can be rewritten as
\begin{equation}
\ybf_{CFO} = \Psibf \sbf + \nbf,
\end{equation}
where $\Psibf =[\Psibf_1\Fbf,\cdots,\Psibf_K\Fbf]$ and $\Psibf_k = \Dbf_k \Lambdabf_k\Pbf_k$.

We assume that full channel state information (CSI) is available at both the transmitters\footnote{The required CSI at the transmitters (CSIT) is the frequency-domain channel response values of $\Lambdabf_k$ corresponding to the nonzero part of $\Pbf_k$ from each user. In most transmit signal design to exploit the channel, CSIT is assumed, e.g., in MIMO precoding. CSI at the receiver (CSIR) is anyway necessary for coherent decoding. One way to resolve the CSIT issue is that the receiver with CSIR designs the GFDM filter and feedbacks the GFDM filter information to the transmitters as in MIMO precoding.} and the receiver and the ZF receiver is used,  and consider two cases regarding the knowledge of the CFO values   $\{\epsilon_1,\cdots,\epsilon_K\}$.
First, we assume that the CFO values  are known to the receiver. In this case, perfect ZF processing  can be performed, i.e.,
\begin{equation} \label{eq:CFOrateKnownEps}
\hat{\sbf}= \Psibf^{-1}\ybf_{CFO}=\sbf + \Psibf^{-1}\nbf,
\end{equation}
 and  the corresponding sum rate  is given by
\begin{equation}\label{eq:ZFGFDMrateCFO}
R_{ZF,{CFO}}(\gammabf,\epsilonbf)=\sum_{n=0}^{N-1}\log
\left(1+\dfrac{\sigma_s^2/\sigma_n^2}{[(\Psibf^H\Psibf)^{-1}]_{n,n}}\right),
\end{equation}
where  $\epsilonbf=[\epsilon_0,\cdots,\epsilon_{K-1}]$. Note that in this case there is no inter-subcarrier interference  as seen in \eqref{eq:CFOrateKnownEps}, and the sum rate in \eqref{eq:ZFGFDMrateCFO} assumes
that the receiver knows all CFO values
$\epsilonbf=[\epsilon_0,\cdots,\epsilon_{K-1}]$ because
the ZF receiver $\Psibf^{-1}$ requires  the CFO information. Note that $\{\epsilon_1,\cdots,\epsilon_K\}$ are the residual CFO after carrier recovery and hence, identifying the (residual) CFO values $\{\epsilon_1,\cdots,\epsilon_K\}$ at the receiver may not be easy. Thus, a more practical assumption is that  the receiver skips extracting  the exact (residual) CFO
information and simply uses the nominal ZF receiver $\hat{\Psibf}^{-1}$, where $\hat{\Psibf}=[\hat{\Psibf}_1\Fbf,\cdots,\hat{\Psibf}_K\Fbf]$ and $\hat{\Psibf}_k= \Lambdabf_k\Pbf_k$ with the assumption $\Dbf_k=\Ibf$. The nominal ZF receiver output is given by
\begin{equation}
\hat{\sbf}=\hat{\Psibf}^{-1}\Psibf\sbf+\hat{\Psibf}^{-1}\nbf.
\end{equation}
Note that the nominal ZF induces inter-subcarrier interference,  the   SINR of the $n$-th subsymbol of $\hat{\sbf}$ is given by
\[
\mathrm{SINR}_{n}^{\widehat{ZF}}=\dfrac{|[\hat{\Psibf}^{-1}\Psibf]_{n,n}|^2\sigma_s^2}{\sum_{i=0,~i\neq
n }^{N-1}|[\hat{\Psibf}^{-1}\Psibf]_{n,i}|^2\sigma_s^2 +
|[\hat{\Psibf}^{-1}]_{n,n}|^2\sigma_n^2},
\]
and the corresponding sum rate is given by
\begin{equation}\label{eq:ZFGFDMratenominal}
R_{\widehat{ZF},{CFO}}(\gammabf,\epsilonbf)=\sum_{n=0}^{N-1}\log
\left(1+\mathrm{SINR}_{n}^{\widehat{ZF}}\right).
\end{equation}
In this case,  the rate-optimal filter design problem  can be formulated as follows:

\vspace{0.5em}
\begin{problem}[Rate maximization for GFDM with nominal ZF with CFO] \label{prob:rateMaxZFCFOB}
\begin{equation}
\begin{array}{cl}
\underset{\gammabf}{\arg \min}&\mathbb{E}_{\epsilonbf }\{ R_{\widehat{ZF},{CFO}}(\gammabf,\epsilonbf)\}\\
~\mathrm{s.t. }~~ &\sum_{m=0}^{2M-1}|\gamma_m|^2=M.
\end{array}
\end{equation}
where $\mathbb{E}_{\epsilonbf}$ is expectation over $\epsilonbf$ with distribution $p(\epsilonbf)$.
\end{problem}
\vspace{0.5em}

In Problem \ref{prob:rateMaxZFCFOB}, the cost function is the sum
rate averaged over the CFO values since the CFO values are unknown
and it is not desirable to  design the filter for a specific set
of CFO values. For the CFO distribution $p(\epsilonbf)$, we may
consider a proper distribution such as a zero-mean Gaussian or
uniform distribution. A closed-form solution to Problem
\ref{prob:rateMaxZFCFOB} seems hard to obtain, but the problem can
  be solved numerically by approximating $\mathbb{E}_{\epsilonbf }$ with the Monte Carlo method\cite{MonteCarlo:book}. GFDM filter optimization for rate
maximization under the MMSE receiver with CFO can be done in a
similar way. Note that in Problem \ref{prob:rateMaxZFCFOB} the
GFDM filter is optimized to maximize the data rate against CFO. It
will be seen later in Section \ref{sec:NumericalResult} that the
optimized GFDM filter in this way provides a non-trivial rate gain
over SC-FDM especially at high SNR and  the GFDM filtering is
beneficial for the data rate even under linear receivers when the
channel is perturbed by CFO.

\section{Joint Design of GFDM Filter and Window}
\label{sec:jointDesign}
In this section, we consider the joint design of window and GFDM
filter for the two design criteria  in Section
\ref{sec:filterdesignExamples}. For the windowing technique we
consider windowing  based on both prefix and suffix
\cite{Prasad:book,Plass:book}. In this method, the first $N_w (<
N_{cp}$) samples $x_k[0],x_k[1],\cdots,x_k[N_w-1]$ of subcarrier
$k$'s signal $\xbf_k = \Wbf_N^H\Pbf_k \Gammabf \Rbf \Wbf_M \sbf_k$
are attached to the end of the vector $\xbf_k$ to yield a suffix
of size $N_w$, while the last $N_{cp}$ samples
$x_k[N-N_{cp}],\cdots,x_k[N-1]$ of $\xbf_k$ are attached in the
front of $\xbf_k$ just as a usual CP. Then, the first $N_w$
samples of the CP and the last $N_w$ suffix samples are
symmetrically tapered with a proper window
$\wbf^T=[w_1,\cdots,w_{N_w},1,\cdots,1,w_{N_w},\cdots,w_1]$ with $||\wbf||_2=1$, the suffix of the
previous symbol is overlapped with the first $N_w$ samples of the
prefix of the current symbol, and finally the edge-overlapped GFDM
symbols are transmitted.  In this way, no additional time samples
are required for windowing. At the receiver side, by taking only
the $N$ samples $y[0],y[1],\cdots,y[N-1]$ before the suffix part
for each symbol, the same received signal model as in the case of
no windowing can be obtained except the first
$y[0],\cdots,y[N_w-1]$ samples may be corrupted by additive
interference from the suffix part of the previous symbol. (If the
possibility of the interference from the previous symbol needs to
be eliminated completely, a CP of size $N_w+N_{cp}$ samples may be
used with sacrificing the rate a bit.)

Now, consider the problem of rate maximization under CFO. If the
aforementioned  windowing and receiver-sampling  technique is
used, the inner received signal model is untouched. Hence,  the
rate optimization result in Section \ref{subsec:CFOrobustness}
does not change, and the GFDM filter design for rate maximization
under CFO  and the window design  can be separated. In the case of
rate maximization under CFO,  there is a trade-off between OOB
radiation reduction  and rate enhancement for the GFDM filter
 itself since the rate-optimal filter is not the OOB-optimal one.  So, one reasonable approach in this case is that we use
the rate-optimal GFDM filter against CFO
perturbation proposed in Section \ref{subsec:CFOrobustness} to maximize the rate and use the window to simultaneously
reduce the OOB radiation without sacrificing  the
rate. Here, we can adopt one of many windowing functions proposed for OOB reduction  \cite{Prasad:book,Plass:book}.

On the other hand, in the case of OOB reduction, the design of GFDM filter and window is intertwined. In the case of windowing with a prefix of $N_{cp}$ samples and a suffix of $N_w$ samples, the overall transmit signal is cyclo-stationary with period $T_{cp}+T_b$ and the constituent signal given by
\begin{equation}  \label{eq:CTsignalCP}
x_{T_b^\prime}(t)=\sum_{k=0}^{K-1}
e^{\iota2\pi kt/T_s}\sum_{m=0}^{M-1} w(t) {\bar{g}_m(t)}s_{kM+m}, %
\end{equation}
where $ \bar{g}_m(t)=\frac{1}{\sqrt{NM}} \sum_{q=0}^{M-1}e^{-\iota
2\pi mq/M} e^{ \iota 2\pi qt/(MT_s)}(\gamma_q+e^{-\iota 2\pi
t/T_s}\gamma_{M+q})$ for $-T_{cp}\leq t \leq T_b+T_w$ and
$\bar{g}_m(t)=0$ for $ t \notin [-T_{cp}, T_b+T_w]$, and $w(t) \ne
0$ only for $t \in [-T_{cp}, T_b+T_w]$ is the window function.
Here, $T_{cp}$ and $T_w$ are the time intervals for $N_{cp}$ and
$N_w$ samples, respectively. The baseband PSD of this
cyclostationary signal is given by  \eqref{eq:PBBf} with $G_m(f)$
replaced with $W(f) * \bar{G}_m(f)$, where $W(f)$ and
$\bar{G}_m(f)$ are the Fourier transforms of $w(t)$ and
$\bar{g}_m(f)$, respectively, and $*$ represents convolution. The
overall PSD $P(f)$ is obtained by $K$ shifted sum of the baseband spectra
as in \eqref{eq:PBBfall}. One suboptimal way to the joint design
of GFDM filter and window for OOB emission minimization under the
assumption of the MF/SIC receiver is that we simply design the
GFDM filter to minimize OOB emission by solving Problem
\ref{prob:oobmfsic} with the modified PSD including the prefix and
the suffix and additionally apply a known window function.
However, this design does not have any optimality. To solve the
joint optimization problem for OOB radiation minimization,  we can
adopt an alternating optimization technique widely used for joint
optimization when direct joint optimization is difficult
\cite{Csiszar&Tusnady:84}. Applying  this method, we propose the following algorithm for the joint design:

\begin{algorithm} \label{algo:OOBjoint} Initialize the window function $\wbf$ properly.
First,
optimize the GFDM filter as
\begin{equation}  \label{eq:jointgamma}
\underset{\gammabf}{
\min}\underset{f\in[\frac{1}{T_s},\infty)}{\max}
P(-\frac{1}{T_s}-f)+P(\frac{K}{T_s}+f)
\end{equation}
under $\sum_{m=0}^{2M-1}|\gamma_m|^2=M$ for the given window
function. Then, for the obtained GFDM
filter from \eqref{eq:jointgamma},  optimize the
window function as
\begin{equation} \label{eq:jointwin}
\underset{w_1,\cdots,w_{N_w}}{
\min}\underset{f\in[\frac{L}{T_s},\infty)}{\max}
P(-\frac{1}{T_s}-f)+P(\frac{K}{T_s}+f)
\end{equation}
under $||[w_1,\cdots,w_{N_w},1,\cdots,1,w_{N_w},\cdots,w_1]||_2=1$. Iterate this procedure until convergence.
\end{algorithm}

\vspace{0.3em}

 Note that   we introduced the parameter $L \ge 1$ in \eqref{eq:jointwin}.  This is because typically windowing changes the slope of  the spectral skirt of MC signals. Hence, we target enlarging the slope of the spectral decay by choosing the stop band  away from the in-band with some large $L$.
Although
Algorithm \ref{algo:OOBjoint} does not guarantee global optimality, it monotonically improves the
performance at each iteration and a local optimality is
guaranteed. The result of the joint design is provided in Section \ref{sec:NumericalResult}.

\section{Numerical results}
\label{sec:NumericalResult}


\subsection{OOB Radiation Minimization}
\label{sec:numOOB}

\begin{figure}[http]
    \centerline{\scalefig{0.47}\epsfbox{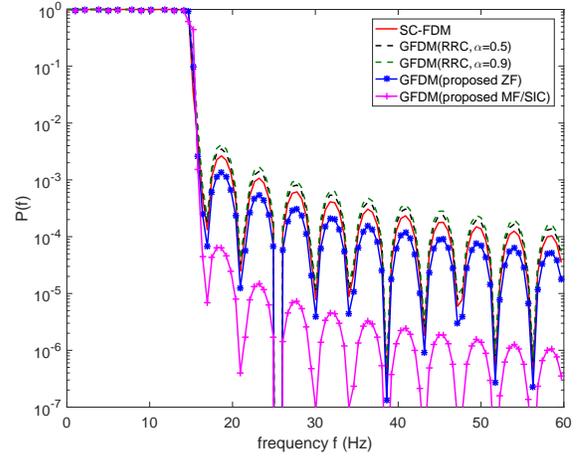}}
    \captionsetup{justification=centering}
\caption{OOB radiation: $K=30$, $M=9$, $T_s=1$, $\sigma_n^2=1$, and $P_s=1$} \label{fig:OOBnoCFO}
\end{figure}

We evaluated the GFDM filter design method for OOB radiation
minimization presented in Section \ref{sec:filterdesignother}. The
considered GFDM  parameters are $K=30$, $M=9$ ($N=KM=270$),
$\sigma_n^2=1$, and one subsymbol duration is normalized to be
$T_s=1$ sec (one GFDM symbol interval $T_b$ is $MT_s$).   First,
we considered the case of no windowing with no CP and no suffix.
We solved Problems \ref{prob:oobmfsic} and
 \ref{prob:oobZF} with $P_s=1$ to obtain optimal GFDM filters for OOB radiation minimization in  MF/SIC and ZF receiver cases, respectively, with the rates computed under the assumption of AWGN. The result is shown in Fig. \ref{fig:OOBnoCFO}. For comparison, the PSD curves of SC-FDM and GFDM with  RRC filters of roll-off factors $\alpha=0.5$ and $\alpha=0.9$ are also included in Fig.  \ref{fig:OOBnoCFO}.
It is seen that the OOB-optimal GFDM filter targeting OOB
radiation minimization without any constraint except the power
constraint obtained by solving Problem \ref{prob:oobmfsic}
drastically reduces OOB radiation  compared to SC-FDM, and the
GFDM filter obtained by solving Problem \ref{prob:oobZF} under the
assumption of the ZF receiver with the 10 \% rate loss constraint
compared to SC-FDM (i.e., $\eta=0.1$ in \eqref{eq:OOBminZF})
improves the OOB radiation performance   compared to SC-FDM.
However, in the latter case, the OOB radiation improvement is not
so drastic. It is also seen that GFDM with RRC filters with
roll-off factors 0.5 and 0.9 yields slightly worse OOB performance
than SC-FDM.  Note in Fig. \ref{fig:OOBnoCFO} that the decay rate
of OOB radiation  with respect to the frequency seems to be the
same for all the considered GFDM filters including the OOB-optimal
one obtained by solving Problem \ref{prob:oobmfsic}.

\begin{figure}[http]
    \centerline{\scalefig{0.47}\epsfbox{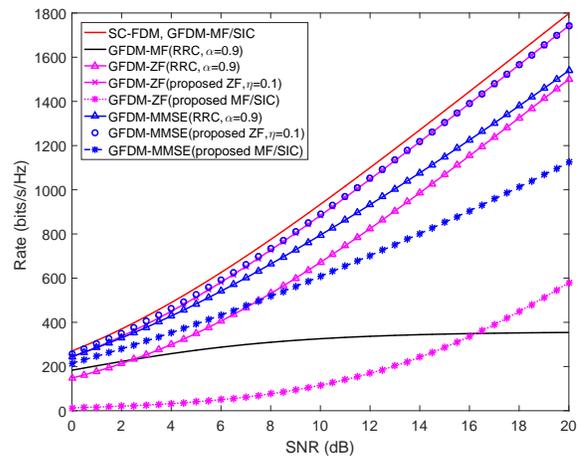}}
    \captionsetup{justification=centering}
\caption{Rate versus SNR}
\label{fig:ratelossComp}
\end{figure}

Fig. \ref{fig:ratelossComp} shows the rate performance
corresponding to the result in Fig. \ref{fig:OOBnoCFO} with the
same system setup and SNR defined as $P_s/\sigma_n^2$. The GFDM
filters were designed at 0 dB SNR (i.e., $P_s=\sigma_n^2=1$) but
the SNR value was swept for the given GFDM filters. First, note
that the rate performance of the GFDM filter obtained by solving
Problem \ref{prob:oobmfsic} assuming the MF/SIC receiver yields
very poor rate performance when the corresponding GFDM filter is
combined with the linear ZF or MMSE receiver. Hence, in this case,
SIC must be used for the desired rate performance. (When combined
with the MF/SIC receiver, it achieves the rate of SC-FDM as
discussed in Section \ref{subsec:MFreceiver}.) In the case of the
GFDM filter obtained by solving Problem \ref{prob:oobZF} with
$\eta=0.1$ at 0 dB SNR under the assumption of the ZF  receiver,
there exists  slight loss in the  rate performance compared to
SC-FDM. Hence, in the linear receiver case, indeed there exists a
trade-off between the rate performance and the OOB radiation
performance for the optimally designed GFDM filter. It is
interesting to note that GFDM with RRC filters is good neither for
the OOB performance nor for the rate performance compared to
SC-FDM. Hence, the RRC filter is  not optimal in terms of rate-OOB
emission trade-off. Note also that the rate performance of GFDM
with a  nontrival RRC filter and the MF receiver  saturates
quickly as SNR increases. Hence, the MF receiver is  not a viable
option for GFDM.

\begin{figure}[http]
    \centerline{\scalefig{0.47}\epsfbox{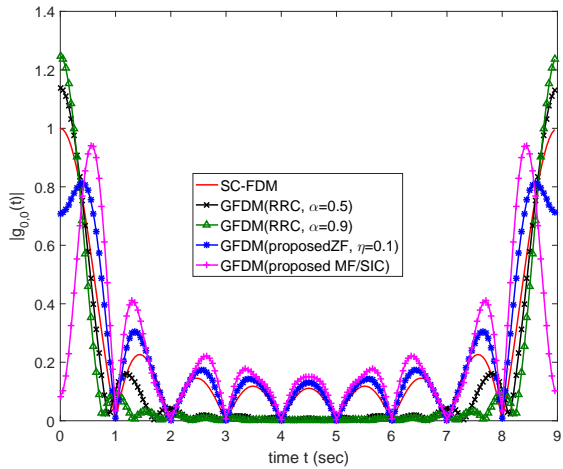}}
    \captionsetup{justification=centering}
\caption{Designed filter response in time domain}
\label{fig:FilRes}
\end{figure}

Fig. \ref{fig:FilRes} shows the time-domain
 filter magnitude response corresponding to  Figs. \ref{fig:OOBnoCFO} and \ref{fig:ratelossComp}.
Note that the
rate-optimal SC-FDM has one major peak,  the OOB-optimal filter obtained by solving Problem \ref{prob:oobmfsic}
has two major peaks, and the filter obtained by solving Problem
\ref{prob:oobZF} seems a mixture of the two filter responses.

\begin{figure}[http]
    \centerline{\scalefig{0.47}\epsfbox{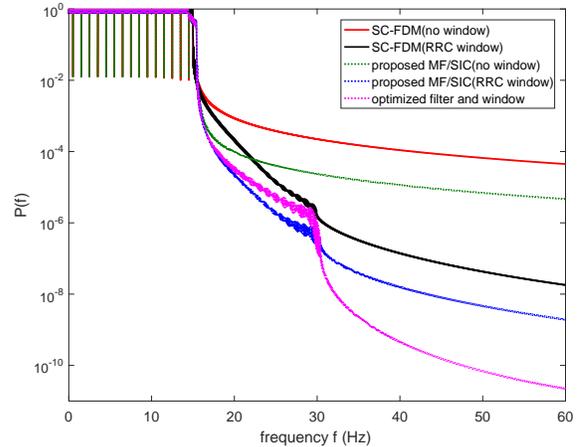}}
    \captionsetup{justification=centering}
\caption{OOB radiation: $K=30$, $M=9$, $T_s=1$, $\sigma_n^2=1$, $P_s=1$, $N_{cp}=30$ (1/9=11.1\%), and $N_w=3$}
\label{fig:OOBwinFil}
\end{figure}

Next, we considered the joint design of GFDM filter and window for
OOB radiation minimization under the assumption of SIC at the receiver. To include windowing, we added
$N_{cp}=30$ samples in the front and $N_w=3$ samples at the end of
each GFDM symbol with the same parameters considered for Fig.
\ref{fig:OOBnoCFO},
 and used the windowing technique based on both prefix and suffix explained in Section  \ref{sec:jointDesign}. We tried the two design methods mentioned in Section \ref{sec:jointDesign}. The first method is that we simply optimized the GFDM filter for OOB emission reduction  by solving Problem  \ref{prob:oobmfsic} with the modified PSD including the prefix and the suffix, and applied a known window function, here a RRC window. The second method is the joint optimization based on Algorithm \ref{algo:OOBjoint}. For Algorithm \ref{algo:OOBjoint}, we initialized the window with a rectangular window and chose $L=30$.
 The result is shown in Fig. \ref{fig:OOBwinFil}. It is seen that with inclusion of  prefix and  suffix the gap between the proposed design and SC-FDM is reduced compared to Fig. \ref{fig:OOBnoCFO} but there still exists significant improvement. It is also seen that with windowing additional OOB emission reduction is achieved, and the reduction is far larger for the joint optimal design of GFDM filter and window than for the suboptimal method for $f \ge 30$.

\subsection{Rate Maximization under CFO}
\label{sec:numRate}

\begin{figure}[ttp]
    \centerline{\scalefig{0.47}\epsfbox{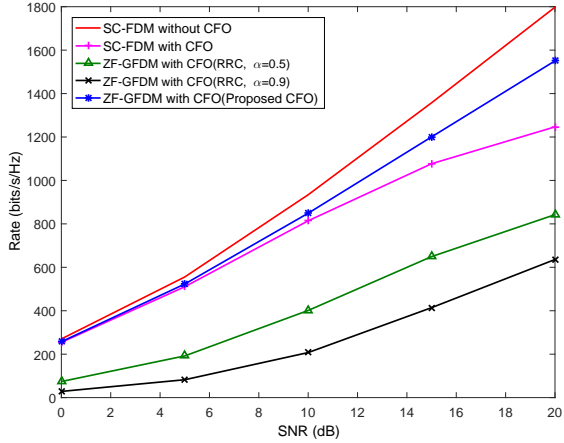}}
    \captionsetup{justification=centering}
\caption{Rate versus SNR: AWGN channel with CFO}
\label{fig:CFOrateAWGN}
\end{figure}
\begin{figure}[ttp]
    \centerline{\scalefig{0.47}\epsfbox{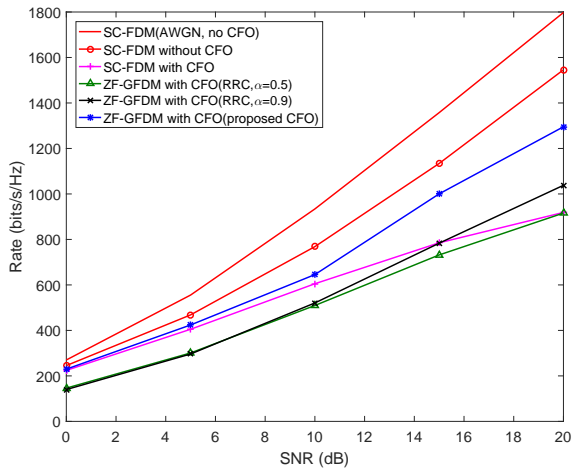}}
    \captionsetup{justification=centering}
\caption{Rate versus SNR: An LTE PB channel with CFO}
\label{fig:CFOratePB}
\end{figure}
\begin{figure}[http]
    \centerline{\scalefig{0.47}\epsfbox{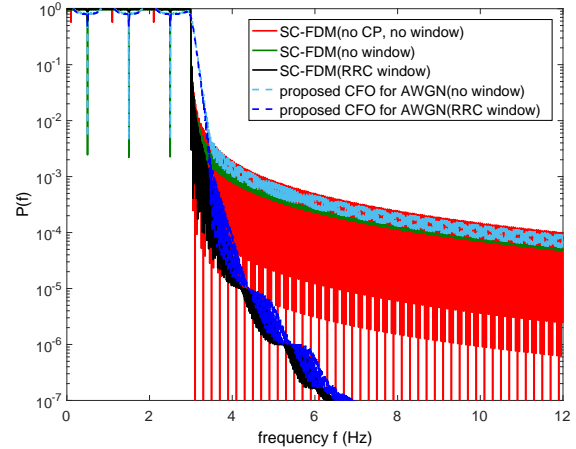}}
    \captionsetup{justification=centering}
\caption{Power spectral density}
\label{fig:OOBPB}
\end{figure}

We considered an uplink GFDM network of $K=6$ users with
$M=45$. The CFO $\epsilon_k$ for user $k$ was generated according
to the uniform distribution $\epsilon_k \sim
\mbox{Unif}[-1/200,1/200]=[-0.5\%,0.5\%]$. We considered two channels with CFO:
AWGN and the Pedestrian B (PB) channel with six non-zero multiple
paths of LTE \cite{3GPPstandard}. We considered the SNR range from
0 to 20 dB. To obtain a rate-optimal filter against CFO, we solved
Problem \ref{prob:rateMaxZFCFOB} for each of  the SNR values  0,
5, 10, 15, and 20 dB. The results are shown in Figs.
\ref{fig:CFOrateAWGN} and \ref{fig:CFOratePB}.  It is seen in Fig.
\ref{fig:CFOrateAWGN} that in the AWGN with CFO, the performance
of SC-FDM degrades from the case of no CFO and the proposed GFDM
filter outperforms SC-FDM. It is also seen that the performance of
the RRC filter is bad compared to other filters.  In the general
PB channel case, similar trends are seen  in Fig.
\ref{fig:CFOratePB} In both cases, the performance gain of the
proposed GFDM filter is large compared to SC-FDM at high SNR. This
is because at high SNR  the dominant factor for performance
degradation is not noise but CFO. Hence, handling CFO properly
with a well designed GFDM filter yields larger gain at high SNR.
Finally, we compared the OOB
radiation of the GFDM filter used for Fig. \ref{fig:CFOrateAWGN},
and the result is shown in Fig. \ref{fig:OOBPB}. It is seen that
in the case of no window the OOB radiation decaying slope with
respect to the frequency seems the same for SC-FDM and the
proposed filter. In the windowing case, the spectrum skirt
decreases significantly for both SC-FDM and the proposed GFDM
filter used for Fig. \ref{fig:CFOrateAWGN}. Note that with the
normalized $T_s=1$, the subcarrier spacing is 1 Hz. The center
frequencies  of the six subcarriers ($K=6$) are -2.5, -1.5, -0.5,
0.5, 1.5, and 2.5 Hz. In the case of SC-FDM, the waveform at each
subcarrier occupies 1 Hz. In the case of GFDM with the repetition
factor $L=2$, on the other hand, the waveform at each subcarrier
occupies 2 Hz. So, it is seen that the PSD of SC-FDM suddenly
drops down at 3 (=2.5+1/2) Hz and the PSD of GFDM with $L=2$
extends to 3.5 (=2.5+2/2) Hz. However, with windowing the window
PSD envelope is dominant in both cases and the PSDs of both cases
are very close beyond 3.5 Hz.

\section{Conclusion}
\label{sec:conclusion}

We have considered optimal filter design for
GFDM under the criteria of rate maximization and OOB emission minimization.  In the case of rate maximization, we have shown that the Dirichlet filter is optimal in the AWGN channel with no CFO under linear
receivers, but a properly designed non-trivial GFDM filter can yield better
performance than the Dirichlet filter in general channels perturbed by CFO. In the case of OOB emission minimization, we have shown that drastic OOB reduction is possible by designing the GFDM filter properly under the assumption of MF/SIC receivers. Finally, we have shown that windowing can further enhance the performance when combined with a properly designed GFDM filter.

\section*{Appendix} \label{sec:appendix1}

{\em Proof of Theorem \ref{lem:MFGFDMrate}:}  ~ Note that
\begin{equation}
{\Pbf}_k^T {\Pbf}_i=\left\{\begin{array}{cl}\Ibf_{2M} & \textrm{ if } k=i~ (\mathrm{mod} ~K),\\
\left[\begin{array}{cc}\mathbf{0}&\Ibf_{M}\\\mathbf{0}&\mathbf{0}\end{array}\right]&\textrm{ if } k=i-1~ (\mathrm{mod} ~K),\\
\left[\begin{array}{cc}\mathbf{0}&\mathbf{0}\\\Ibf_{M}&\mathbf{0}\end{array}\right]&\textrm{ if } k=i+1~ (\mathrm{mod} ~K),\end{array}\right.
\end{equation}
for $i=0,\cdots,K-1$.  Then,
\begin{align}
&\Fbf^H{\Pbf}_k^T {\Pbf}_i\Fbf=\Wbf_M^H\Rbf^T \Gammabf^* \Pbf_i^T\Pbf_k \Gammabf\Rbf\Wbf_M  \label{eq:appendTh1GPkT}\\
&=\left\{\begin{array}{cl}\Wbf_M^H(\Gammabf^{(f)*}\Gammabf^{(f)}+\Gammabf^{(r)*}\Gammabf^{(r)})\Wbf_M&\textrm{if }i=k,\\
\Wbf_M^H(\Gammabf^{(r)*}\Gammabf^{(f)})\Wbf_M&\textrm{if }i=k-1, \\
\Wbf_M^H(\Gammabf^{(f)*}\Gammabf^{(r)})\Wbf_M&\textrm{if }i=k+1, \\
\mathbf{0}&\textrm{otherwise,}\end{array}\right. \nonumber
\end{align}
where the mod notation is omitted.
Then, from \eqref{eq:MF} we have $\hat{\sbf}_k=\underbrace{\Wbf_M^H(\Gammabf^{(f)*}\Gammabf^{(f)} +\Gammabf^{(r)*}\Gammabf^{(r)})\Wbf_M}_{\Abf_k}\sbf_k
+\underbrace{\Wbf_M^H(\Gammabf^{(r)*}\Gammabf^{(f)})\Wbf_M}_{\Abf_{-k}}\sbf_{k-1}+\underbrace{\Wbf_M^H(\Gammabf^{(f)*}\Gammabf^{(r)})\Wbf_M}_{\Abf_{+k}}\sbf_{k+1}+\Fbf^H\Pbf_k^T\Wbf_N\nbf.$
Note that  $\tilde{\nbf}= \Wbf_N\nbf \sim \mathcal{CN}(0,\sigma_n^2\Ibf_N)$ since the DFT matrix $\Wbf_N$ unitary and
\begin{align} \label{eq:appendTheo1noisePower}
\Fbf^H \Pbf_k^T \tilde{\nbf} &\sim\mathcal{CN}(\mathbf{0},\Wbf_M(\Gammabf^{(f)*}\Gammabf^{(f)}+\Gammabf^{(r)*}\Gammabf^{(r)})\Wbf_M^H)
\end{align}
$\forall~k$ from \eqref{eq:appendTh1GPkT}.
Hence, the estimated signal of the $m$-th subsymbol of the $k$-th subcarrier is given by $\hat{s}_{kM+m}=\sum_{p=0}^{M-1}[\Abf_k]_{m,p}s_{kM+p}+[\Abf_{-k}]_{m,p}s_{(k-1)M+p}+[\Abf_{+k}]_{m,p}s_{(k+1)M+p}+[\Fbf^H \Pbf_k^T \tilde{\nbf}]_m.$ With some computation, we have
\begin{align}
\hat{s}_{kM+m}=\frac{1}{M}\sum_{p=0}^{M-1}\sum_{l=0}^{M-1}e^{\iota 2 \pi (m-p)l/M} \bigg[ (|\gamma_l|^2+|\gamma_{M+l}|^2)s_{kM+p}\nonumber\\
+(\gamma_{M+l}^*\gamma_l)s_{(k-1)M+p}+(\gamma_l^*\gamma_{M+l})s_{(k+1)M+p}\bigg]
+[\Fbf^H \Pbf_k^T \nbf']_m. \label{eq:estMF}
\end{align}
In \eqref{eq:estMF}, $s_{kM+m}$ is the signal, $s_{kM+p},~ p\neq m$ is inter-subsymbol/inter-subcarrier interference, and $[\Fbf^H \Pbf_k^T \tilde{\nbf}]_m$ is noise. The interference power is  $a(\gammabf)P_s$ and the noise power is $\frac{\sum_{l=0}^{M-1}(|\gamma_l|^2+|\gamma_{M+l}|^2)}{M}\sigma_n^2=\sigma_n^2$ from \eqref{eq:appendTheo1noisePower} and the filter power constraint. Hence, the MF SINR is given by $\frac{P_s}{a(\gammabf)P_s+\sigma_n^2}$.
\hfill $\blacksquare$

\vspace{1em}

{\em Proof of Theorem \ref{lem:ZFGFDMrate}:} ~  From \eqref{eq:ZFprocess}
 the  ZF equalizer after DFT is given by $\Bbf:=(\Pbf\mathrm{diag}(\Fbf,\cdots,\Fbf))^{-1}\Lambdabf_H^{-1}$ and hence the output SNR of the $m$-th subsymbol of the $k$-th subcarrier is given by \cite{JiangVaranasiJianLi:11IT}
\begin{equation}\label{eq:effective}
\rho_{k,m,ZF}(\gammabf)=\dfrac{P_s/\sigma_n^2}{[\Bbf\Bbf^H]_{kM+m,kM+m}}.
\end{equation}
The matrix  $\Pbf\mathrm{diag}(\Fbf,\cdots,\Fbf)=[\Pbf_0\Fbf, \cdots, \Pbf_{K-1}\Fbf]$ in $\Bbf$ can be written after some computation as
\begin{align}\label{eq:appendPdiagG}
&\left[\begin{array}{cccc}\Gammabf^{(f)}&\Gammabf^{(r)}&\cdots&\mathbf{0}\\
\mathbf{0}&\Gammabf^{(f)}&\ddots&\vdots\\
\vdots&\ddots&\ddots&\Gammabf^{(r)}\\
\Gammabf^{(r)}&\cdots&\cdots&\Gammabf^{(f)}\end{array}\right]\mathrm{diag}(\Wbf_M ,\cdots, \Wbf_M)
\end{align}
Then, $\Bbf=\Wbf\Cbf\Lambdabf_{H}^{-1}$, where
 $\Wbf:=\mathrm{diag}(\Wbf_M^H,\cdots,\Wbf_M^H)$ and
\begin{equation} \label{eq:appendCbf}
\Cbf:=\left[\begin{array}{cccc}\Gammabf^{(f)}&\Gammabf^{(r)}&\cdots&\mathbf{0}\\
\mathbf{0}&\Gammabf^{(f)}&\ddots&\vdots\\
\vdots&\ddots&\ddots&\Gammabf^{(r)}\\
\Gammabf^{(r)}&\cdots&\cdots&\Gammabf^{(f)}\end{array}\right]^{-1}.
\end{equation}
From the fact that $\Wbf$ is a block diagonal matrix of repeated $\Wbf_M^H$, the diagonal elements of $\Bbf\Bbf^H$ is given by
\begin{align}
 [\Bbf\Bbf^H]_{kM+m,kM+m}&=\sum_{i=0}^{N-1} |[\Bbf]_{kM+m,i}|^2\nonumber\\
&=\sum_{i=0}^{N-1}|[\Wbf\Cbf\Lambdabf_H^{-1}]_{kM+m,i}|^2\nonumber\\
&=\sum_{i=0}^{N-1}\left|\sum_{j=0}^{N-1}[\Wbf]_{kM+m,j}[\Cbf]_{j,i}[\Lambdabf_{H}^{-1}]_{i,i}\right|^2\nonumber\\
&=\sum_{i=0}^{N-1}\left|\sum_{t=0}^{M-1}\frac{e^{\iota 2\pi mt/M}}{\sqrt{M}}[\Cbf]_{kM+t,i}[\Lambdabf_{H}^{-1}]_{i,i}\right|^2.\label{eq:BBzf}
\end{align}
Using the fact that a block circulant matrix can be block-diagonalized by block DFT  \cite{Mazancourt:83TAP}, we have
\begin{equation}\label{eq:appendCinverse}
\Cbf^{-1}=\left[\begin{array}{cccc}\Gammabf^{(f)}&\Gammabf^{(r)}&\cdots&\mathbf{0}\\
\mathbf{0}&\Gammabf^{(f)}&\ddots&\vdots\\
\vdots&\ddots&\ddots&\Gammabf^{(r)}\\
\Gammabf^{(r)}&\cdots&\cdots&\Gammabf^{(f)}\end{array}\right]=\Ubf\Dbf\Ubf^{H},
\end{equation}
where $\Ubf$ is  the unitary block DFT matrix  given by
\begin{equation}  \label{eq:appendZFUbf}
\Ubf=\dfrac{1}{\sqrt{K}}\left[\begin{array}{cccc}\Ibf&\Ibf&\cdots&\Ibf\\\Ibf&e^{\iota2\pi/K}\Ibf&\cdots &e^{\iota2\pi(K-1)/K}\Ibf\\
\vdots&\ddots&\ddots&\vdots\\\Ibf&e^{\iota2\pi(K-1)/K}\Ibf&\cdots&e^{\iota2\pi(K-1)^2/K}\Ibf\end{array}\right],
\end{equation}
and
\begin{equation} \label{eq:appendZFBigDbf}
\Dbf=\left[
\begin{array}{cccc}
\dbf_0&\mathbf{0}&\cdots&\mathbf{0}\\
\mathbf{0}&\dbf_1 &\cdots&\mathbf{0}\\
\vdots&\ddots&\ddots&\vdots\\
\mathbf{0}&\cdots&\cdots&\dbf_{K-1}
\end{array}
\right]
\end{equation}
with $M\times M$ diagonal matrices
\begin{equation} \label{eq:appendZFdbfl}
\dbf_{l}=\Gammabf^{(f)}+e^{\iota2\pi l /K }\Gammabf^{(r)},~l=0,\cdots,K-1.
\end{equation}
 Hence, $\Cbf=(\Ubf\Dbf\Ubf^H)^{-1}=\Ubf\Dbf^{-1}\Ubf^H$.
By changing the variable $i$ in \eqref{eq:BBzf} to $i=pM+q$, $p=0,\cdots,K-1,~q=0,\cdots,M-1$, we have  $[\Cbf]_{kM+t,i}$  in \eqref{eq:BBzf} as
\begin{align}\label{eq:Cbftq}
[\Cbf]_{kM+t,pM+q}&=[\Ubf \Dbf^{-1}\Ubf^H]_{kM+t,pM+q}\nonumber\\
&=[\Ubf(k,:) \Dbf^{-1} \Ubf(p,:)^H]_{t,q}\nonumber\\
&=\dfrac{1}{K} \sum_{l=0}^{K-1} e^{\iota2\pi(k-p)l/K} [\dbf_l^{-1}]_{t,q}\nonumber\\
&=\delta_{t,q}\dfrac{1}{K} \sum_{l=0}^{K-1} \dfrac{e^{\iota2\pi(k-p)l/K}}{\gamma_{q}+e^{\iota2\pi l/K }\gamma_{M+q}},
\end{align}
where $\Ubf(k,:)$ denotes the $k$-th block row of $\Ubf$.
By substituting \eqref{eq:Cbftq} into \eqref{eq:BBzf}, we have
\begin{equation} \label{eq:BB2}
[\Bbf\Bbf^H]_{kM+m,kM+m}=\sum_{p=0}^{K-1} \sum_{q=0}^{M-1}\left|\sum_{l=0}^{K-1} \dfrac{c_{k,p,q,l}}{\gamma_{q}+d_l\gamma_{M+q}} \right|^2,
\end{equation}
where $c_{k,p,q,l}$ and $d_l$ are defined in Theorem \ref{lem:ZFGFDMrate}.
The sum rate of GFDM-ZF is simply the sum of the $N$ parallel Gaussian channels given by
\begin{equation}\label{eq:effectiveCap}
R_{ZF}(\gammabf)=\sum_{k=0}^{K-1}\sum_{m=0}^{M-1}\log\left(1+\rho_{k,m,ZF}(\gammabf)\right).
\end{equation}
Combining \eqref{eq:effective},
\eqref{eq:BB2} and  \eqref{eq:effectiveCap} yields the desired result
 \eqref{eq:ZFcap}.

In the AWGN channel case of $\Lambdabf_H=\Ibf_N$, \eqref{eq:BB2} becomes \eqref{eq:BBAWGN}, where step (a) is simple  change of the summation order and step (b) is valid since $\sum_{p=0}^{K-1}e^{\iota2\pi(k-p)(l-s)/K}=K$ if $l=s$ and $\sum_{p=0}^{K-1}e^{\iota2\pi(k-p)(l-s)/K}=0$ if $l\ne s$.
\begin{figure*}
\begin{align}
[\Bbf\Bbf^H]_{kM+m,kM+m}
&=\sum_{p=0}^{K-1} \sum_{q=0}^{M-1}\left|\dfrac{1}{\sqrt{M}}\dfrac{1}{K}\sum_{l=0}^{K-1} \dfrac{e^{\iota2\pi(k-p)l/K}}{\gamma_{q}+d_l\gamma_{M+q}}\right|^2\nonumber\\
&=\frac{1}{MK^2}\sum_{p=0}^{K-1} \sum_{q=0}^{M-1}\sum_{l=0}^{K-1}\sum_{s=0}^{K-1}  \dfrac{e^{\iota2\pi(k-p)(l-s)/K}}{(\gamma_{q}+d_l\gamma_{M+q})(\gamma_{q}^*+d_s^*\gamma_{M+q}^*)}\nonumber\\
&\stackrel{(a)}{=}\frac{1}{MK^2}\sum_{q=0}^{M-1}\sum_{l=0}^{K-1}\sum_{s=0}^{K-1}\dfrac{1}{(\gamma_{q}+d_l\gamma_{M+q})(\gamma_{q}^*+d_s^*\gamma_{M+q}^*)}
\left(\sum_{p=0}^{K-1}e^{\iota2\pi(k-p)(l-s)/K}\right)\nonumber\\
&\stackrel{(b)}{=}\frac{1}{MK}\sum_{q=0}^{M-1}\sum_{l=0}^{K-1}\left|\dfrac{1}{\gamma_{q}+d_l\gamma_{M+q}}\right|^2. \label{eq:BBAWGN}
\end{align}
\end{figure*}
Hence, the desired result \eqref{eq:ZFawgnrate} follows.
\hfill $\blacksquare$

\vspace{1em}
\begin{remark}[Low-Complexity Computation of ZF
Equalization] \label{remark:appendLowComplex} For ZF equalization,
$\hat{\sbf}=(\Pbf\mathrm{diag}(\Fbf,\cdots,\Fbf))^{-1}\Lambdabf_H^{-1}\Wbf_N\ybf$
should be computed as seen in \eqref{eq:ZFprocess}.  The left
multiplication by $\Wbf_N$ is simple $N$-point DFT, the left
multiplication by $\Lambdabf_H^{-1}$ is elementwise scaling since
$\Lambdabf_H$ is diagonal, and from \eqref{eq:appendPdiagG},
\eqref{eq:appendCbf} and \eqref{eq:appendCinverse}
$(\Pbf\mathrm{diag}(\Fbf,\cdots,\Fbf))^{-1}=\Wbf\Cbf=\mathrm{diag}(\Wbf_M^H,\cdots,\Wbf_M^H)\Ubf\Dbf^{-1}\Ubf^H$.
Hence, the last step can easily be implemented by sequentially
applying block IDFT, elementwise scaling, block DFT and finally
$K$ separate $M$-point IDFTs.
\end{remark}

\vspace{1em}

{\em Proof of Theorem \ref{theo:DiriOpt}:}  ~ First, note that the Dirichlet filter $\gamma_0=\cdots=\gamma_{M-1}= 1$ and
$\gamma_{M}=\cdots=\gamma_{2M-1}=0$) satisfies the power constraint $\sum_{m=0}^{2M-1} |\gamma_m|^2=M$.  From \eqref{eq:ZFawgnrate}, the corresponding rate is given by $N\log(1+P_s/\sigma_n^2)$ because $\sum_{l=0}^{K-1}\sum_{q=0}^{M-1}1/|\gamma_q + \gamma_{M+q}|^2=N$ for the Dirichlet filter. Next, we  show      that $N$ is the minimum value
of the cost function $\sum_{l=0}^{K-1} \sum_{q=0}^{M-1}\dfrac{1}{|\gamma_{q}+d_l\gamma_{M+q}|^2}$ for all filters satisfying the power
constraint. For this, we express the power constraint as  $\sum_{q=0}^{M-1} g_q = M$, where
$g_q :=|\gamma_q|^2+|\gamma_{M+q}|^2$.
Then, the cost function is rewritten as
\begin{align}
&\sum_{l=0}^{K-1}\sum_{q=0}^{M-1}\dfrac{1}{|\gamma_{q}|^2+|\gamma_{M+q}|^2+2\mathrm{Re}(d_l\gamma_q^*\gamma_{M+q})}\nonumber\\
&=\sum_{q=0}^{M-1}\sum_{l=0}^{K-1}\dfrac{1}{g_q+2\mathrm{Re}(d_l\gamma_q^*\gamma_{M+q})}
\label{eq:ZFopt}
\end{align}
For each given $q$,
$2\mathrm{Re} (d_l\gamma_q^*\gamma_{M+q}) > -g_q$ for all
$l=0,\cdots,K-1$ since
$|\gamma_{q}+d_l\gamma_{M+q}|^2=g_q+2\mathrm{Re}
(d_l\gamma_q^*\gamma_{M+q})>0$. Furthermore, for any
$(\gamma_q,\gamma_{M+q})$ we have $ \frac{1}{K}\sum_{l=0}^{K-1}
2\mathrm{Re} (d_l\gamma_q^*\gamma_{M+q}) =\frac{2}{K}\mathrm{Re}
(\sum_{l=0}^{K-1}
d_l\gamma_q^*\gamma_{M+q})=\frac{2}{K}\mathrm{Re}
(\gamma_q^*\gamma_{M+q}\sum_{l=0}^{K-1}
d_l)=0 $ since $d_0 =1, d_1 =
e^{\iota 2\pi/K},\cdots,d_{K-1}= e^{\iota 2\pi (K-1)/K}$ are
located on the complex unit circle with equal spacing for $K > 1$.
Now applying   Jensen's inequality, we have {\small
\begin{align*}
\dfrac{1}{K}\sum_{l=0}^{K-1}\dfrac{1}{g_q+2\mathrm{Re}(d_l\gamma_q^*\gamma_{M+q})}
&\geq\dfrac{1}{g_q+\dfrac{1}{K}\sum_{l=0}^{K-1}2\mathrm{Re} ( d_l
\gamma_q^* \gamma_{M+q})}\\
&= \frac{1}{g_q}
\end{align*}
}since the function $\dfrac{1}{g_q + x}$ is convex for $x>-g_q$
and $ \frac{1}{K}\sum_{l=0}^{K-1} 2\mathrm{Re}
(d_l\gamma_q^*\gamma_{M+q}) =0 $.
 Thus,
\eqref{eq:ZFopt} is lower bounded as  {\footnotesize
\[
\sum_{q=0}^{M-1}\sum_{l=0}^{K-1}\dfrac{1}{g_q+2\mathrm{Re}(d_l\gamma_q^*\gamma_{M+q})} \ge K\sum_{q=0}^{M-1} \dfrac{1}{g_q} \geq  \dfrac{N}{\dfrac{1}{M}\sum_{q=0}^{M-1} g_q} = N
\]
}by the Jensen's inequality since $1/x$ is convex for $x>0$ and
$g_q>0$ for all $q$. Therefore, the GFDM-ZF rate of any filter
satisfying the power constraint is less than or equal to $N\log
(1+P_s/\sigma_n^2).$
\hfill{$\blacksquare$}

\vspace{1em}

{\em Proof of Theorem \ref{theo:MMSErate}:}  ~  From \eqref{eq:PsibfinModel}, \eqref{eq:appendPdiagG} and \eqref{eq:appendCinverse}, we have $\Phibf=\Wbf_N^H\Cbf^{-1}\Wbf^H$, where $\Cbf$ and $\Wbf$ are defined in Proof of Theorem \ref{lem:ZFGFDMrate}, and
$\Phibf^H\Phibf=\mathrm{diag}(\Wbf_M^H ,\cdots, \Wbf_M^H) \Ubf
\Dbf^H\Dbf\Ubf^H\mathrm{diag}(\Wbf_M ,\cdots, \Wbf_M)$. Let
$\Vbf:=\mathrm{diag}(\Wbf_M^H ,\cdots, \Wbf_M^H) \Ubf$. Then, $\Vbf$ is a
unitary matrix since $\Ubf$ and $\mathrm{diag}(\Wbf_M ,\cdots,
\Wbf_M)$ are unitary. Then,
\begin{align}
\left(\Phibf^H\Phibf+\dfrac{\sigma_n^2}{P_s}\Ibf_N\right)^{-1}&=\left(\Vbf\Dbf^H\Dbf\Vbf^H + \dfrac{\sigma_v^2}{\sigma_s^2}\Ibf_N\right)^{-1}\nonumber\\
&=\Vbf\left(\Dbf^H\Dbf+\dfrac{\sigma_n^2}{P_s}\Ibf_N\right)^{-1}\Vbf^H.\label{eq:V}
\end{align}
Therefore, the $(kM+m)$-th diagonal element of \eqref{eq:V} is
$\sum_{i=0}^{N-1}|[\Vbf]_{kM+m,i}|^2\left[\Dbf^H\Dbf+\dfrac{\sigma_v^2}{P_s}\Ibf\right]_{i,i}^{-1}$.
Due to  \eqref{eq:appendZFBigDbf} and \eqref{eq:appendZFdbfl},  changing  the variable $i$ with $i=lM+q,~l=0,\cdots,K-1,~q=0,\cdots,M-1$, we have
\begin{equation}\label{eq:D}
\left[\Dbf^H\Dbf+\dfrac{\sigma_n^2}{P_s}\Ibf_N\right]_{lM+q,lM+q}=|\gamma_q
+ e^{\iota2\pi l/K} \gamma_{M+q}|^2 +
\dfrac{\sigma_n^2}{P_s},
\end{equation}
and from the definitions of $\Vbf$ and $\Ubf$ (see \eqref{eq:appendZFUbf})
\begin{equation}\label{eq:V2}
|[\Vbf]_{kM+m,lM+q}|^2= \left|\dfrac{1}{\sqrt{MK}}e^{-\iota 2\pi
mq/M} e^{\iota2\pi kl/K}\right|^2=\dfrac{1}{N}.
\end{equation}
Substituting \eqref{eq:D} and \eqref{eq:V2} into
\eqref{eq:effectiveMMSE} together with the fact $R_{MMSE}(\gammabf)=\sum_k\sum_m \log (1 + \rho_{k,m,MMSE}(\gammabf))$ yields the desired result \eqref{eq:MMSEawgnrate}. \hfill$\blacksquare$

\begin{figure*}
\begin{align}
x[n]&=\sum_{p=0}^{K-1}\sum_{q=0}^{M-1}\sum_{k=0}^{K-1}\sum_{m=0}^{M-1}\frac{1}{\sqrt{N}}e^{\iota2\pi n (pM+q)/N}[\Ubf \Dbf \Ubf^H]_{pM+q,kM+q} \mathrm{diag}(\Wbf_M ,\cdots, \Wbf_M)_{kM+q,kM+m}s_{kM+m}\label{eq:appendlast1}\\
&=\sum_{p=0}^{K-1}\sum_{q=0}^{M-1}\sum_{k=0}^{K-1}\sum_{m=0}^{M-1}\frac{1}{\sqrt{N}}e^{\iota2\pi n (pM+q)/N}\frac{1}{K} \sum_{l=0}^{K-1} e^{\iota2\pi(p-k)l/K}(\gamma_q+e^{\iota2\pi l/K}\gamma_{M+q})\frac{1}{\sqrt{M}}e^{-\iota2\pi mq/M} s_{kM+m} \label{eq:appendlast2}\\
&=\frac{1}{\sqrt{N}}\frac{1}{K}\frac{1}{\sqrt{M}}\sum_{k=0}^{K-1}\sum_{m=0}^{M-1}\sum_{q=0}^{M-1}\sum_{p=0}^{K-1}e^{\iota2\pi n (pM+q)/N}e^{-\iota2\pi mq/M}s_{kM+m}\sum_{l=0}^{K-1}(e^{\iota2\pi(p-k)l/K}\gamma_q+e^{\iota2\pi(p-k+1)l/K}\gamma_{M+q})\nonumber\\
&=\frac{1}{\sqrt{N}}\frac{1}{\sqrt{M}}\sum_{k=0}^{K-1}\sum_{m=0}^{M-1}\sum_{q=0}^{M-1}e^{\iota2\pi n (kM+q)/N}e^{-\iota2\pi mq/M}s_{kM+m}(\gamma_q+e^{-\iota2\pi n/K}\gamma_{M+q})\label{eq:appendlast3}\\
&=\sum_{k=0}^{K-1}\sum_{m=0}^{M-1}e^{\iota2\pi
kn/K}\underbrace{\left[\frac{1}{\sqrt{K}}\frac{1}{M}\sum_{q=0}^{M-1}e^{\iota2\pi(n-Km)q/N}(\gamma_q+e^{-\iota2\pi
n/K}\gamma_{M+q})\right]}_{g[(n-mK)_{\mathrm{mod} N}]}s_{kM+m}.
\end{align}
\end{figure*}

\vspace{1em}

{\em Proof of Theorem \ref{theo:DiriOptMMSE}:} ~ Proof is similar to that of Theorem \ref{theo:DiriOpt} in the ZF case.   The Dirichlet filter with $\gamma_0=\cdots=\gamma_{M-1}= 1$ and
$\gamma_{M}=\cdots=\gamma_{2M-1}=0$ satisfying  the power constraint $\sum_{m=0}^{2M-1} |\gamma_m|^2=M$ yields the rate  $N\log(1+P_s/\sigma_n^2)$ because  $D(\gamma)$ in \eqref{eq:MMSEawgnrate} is  $D(\gamma)=\frac{1}{1+\sigma_n^2/P_s}$  for the Dirichlet filter. Showing that $N\log(1+P_s/\sigma_n^2)$ is the best rate for GFDM with MMSE receivers is the same as that in the GFDM-ZF case.
The cost function  in Problem \ref{prob:rateMaximizeMMSE} is given by $\sum_{l=0}^{K-1}\sum_{q=0}^{M-1} \dfrac{1}{|\gamma_q + b_l
\gamma_{M+q}|^2+\sigma_n^2/P_s}$. Applying the same techniques as in Proof of Theorem \ref{theo:DiriOpt} to the convex function $f(x)=\frac{1}{x+\sigma_n^2/P_s}$, we have
\begin{equation}
\sum_{l=0}^{K-1}\sum_{q=0}^{M-1} \dfrac{1}{|\gamma_q + b_l
\gamma_{M+q}|^2+\sigma_n^2/P_s}\geq
\dfrac{N}{1+\sigma_n^2/\sigma_s^2},~ \forall \gammabf.
\end{equation}
Hence, $N\log(1+P_s/\sigma_n^2)$ is the best rate for GFDM with MMSE receivers.
\hfill $\blacksquare$

\vspace{1em}

{\em Derivation of Eq. \eqref{eq:tildegnmkggOOB}:}  ~ From \eqref{eq:PsibfinModel}, \eqref{eq:appendPdiagG} and \eqref{eq:appendCinverse}, we have $\xbf=\Wbf_N^H\Cbf^{-1}\Wbf^H \sbf=\Wbf_N^H (\Ubf\Dbf\Ubf^H)\mbox{diag}(\Wbf_M,\cdots,\Wbf_M)\sbf$, where $\Ubf$ and $\Dbf$ are defined in \eqref{eq:appendCinverse}, \eqref{eq:appendZFUbf},  \eqref{eq:appendZFBigDbf} and \eqref{eq:appendZFdbfl}, and hence   $x[n]$ can be rewritten as \eqref{eq:appendlast1}.
For \eqref{eq:appendlast1} and \eqref{eq:appendlast2}, we changed summation over $0,1,\cdots,N-1$ required for the $n$-th element of the product $\Wbf_N^H (\Ubf\Dbf\Ubf^H)$
to double sum over $p=0,\cdots,K-1$ and $q=0,\cdots,M-1$, and used the fact that $\mbox{diag}(\Wbf_M,\cdots,\Wbf_M)$ is block-diagonal and
 $[\Ubf\Dbf\Ubf^H]_{pM+q,kM+r}=$
\begin{align*}
& \sum_{i=0}^{K-1}\sum_{j=0}^{M-1} [\Ubf]_{pM+q,iM+j} [\Dbf]_{iM+j,iM+j} [\Ubf^H]_{iM+j,kM+r}\\
&= \sum_{i=0}^{K-1}\sum_{j=0}^{M-1} \left[\frac{e^{\iota 2\pi pi/K }}{\sqrt{K}}\Ibf\right]_{q,j} [\Dbf]_{iM+j,iM+j} \left[\frac{e^{-\iota 2\pi ik/K}}{\sqrt{K}}\Ibf\right]_{j,r}\\
&= \sum_{i=0}^{K-1} \frac{e^{\iota 2\pi(p-k)i/K}}{K}[\Dbf]_{iM+q,iM+q}~~~(q=j=r)\\
&= \sum_{i=0}^{K-1} \frac{e^{\iota 2\pi(p-k)i/K}}{K}(\gamma_q+e^{\iota 2\pi i/K}\gamma_{M+q}).
\end{align*}
(For $q\ne r$, $[\Ubf\Dbf\Ubf^H]_{pM+q,kM+r}=0$.)
 For \eqref{eq:appendlast3} we used the facts that $\sum_{l=0}^{K-1}e^{\iota2\pi(p-k)l/K}=K\delta_{p,k}$ and $\sum_{l=0}^{K-1} \e^{\iota2\pi(p-k+1)l/K}=K\delta_{p,k-1}$. \hfill $\blacksquare$

\bibliographystyle{ieeetr}
\bibliography{referenceBibs4}


\end{document}

%% file: macros.tex
\def\psfancypar#1#2{\begingroup\def\par{\endgraf\endgroup\lineskiplimit=0pt}
               \setbox2=\hbox{\large\sc #2}
               \newdimen\tmpht \tmpht \ht2 \advance\tmpht by \baselineskip
               \font\hhuge=Times-Bold at \tmpht
               \setbox1=\hbox{{\hhuge #1}}
               \count7=\tmpht \count8=\ht1
               \divide\count8 by 1000 \divide\count7 by \count8 
               \tmpht=.001\tmpht\multiply\tmpht by \count7 
               \font\hhuge=Times-Bold at \tmpht
               \setbox1=\hbox{{\hhuge #1}}
               \noindent
                \hangindent1.05\wd1
               \hangafter=-2 {\hskip-\hangindent
               \lower1\ht1\hbox{\raise1.0\ht2\copy1}%
                \kern-0\wd1}\copy2\lineskiplimit=-1000pt}

\newcommand{\Phibf}{\mbox{${\bf \Phi}$}}
\newcommand{\Gammabf}{\mbox{${\bf \Gamma}$}}
\newcommand{\Psibf}{\mbox{${\bf \Psi}$}}

\newcommand{\e}{\mbox{${\bf e}_k$}}

\newcommand{\E}{\mbox{{\rm E}}}

\def\boxit#1{\vbox{\hrule\hbox{\vrule\kern3pt
        \vbox{\kern3pt#1\kern3pt}\kern3pt\vrule}\hrule}}

\def\reals{ { {\rm  I \kern-0.15em R }  } }
\def\complex{ {\,{{\rm C} \kern-0.50em \raise0.20ex {  |}}\, }}

\def\gammabf{\hbox{\boldmath$\gamma$\unboldmath}}

\def\epsilonbf{\hbox{\boldmath$\epsilon$\unboldmath}}

\def\mubf{\hbox{\boldmath$\mu$\unboldmath}}

\def\Sigmabf{\hbox{$\bf \Sigma$}}

\def\Omegabf{\hbox{$\bf \Omega$}}

\def\Gammabf{\hbox{$\bf \Gamma$}}

\def\Lambdabf{\mbox{$ \bf \Lambda $}}

\def\Pibf{{\bf \Pi}}

\def\dbf{{\bf d}}

\def\hbf{{\bf h}}

\def\nbf{{\bf n}}

\def\sbf{{\bf s}}

\def\wbf{{\bf w}}
\def\xbf{{\bf x}}
\def\ybf{{\bf y}}

\def\xbf{{\bf x}}
\def\ybf{{\bf y}}
\def\Abf{{\bf A}}
\def\Bbf{{\bf B}}
\def\Cbf{{\bf C}}
\def\Dbf{{\bf D}}

\def\Fbf{{\bf F}}

\def\Hbf{{\bf H}}
\def\Ibf{{\bf I}}

\def\Lbf{{\bf L}}

\def\Pbf{{\bf P}}

\def\Rbf{{\bf R}}

\def\Ubf{{\bf U}}
\def\Vbf{{\bf V}}
\def\Wbf{{\bf W}}

\def\Cc{{\cal C}}

\def\Nc{{\cal N}}

\def\be{\vskip .3cm \begin{equation}}
\def\ee{\end{equation} \vskip .4cm \noindent}
\newcommand{\R}{\mbox{$\hat {\bf R}_{N}$}}

\def\Rxx{\Rbf_{\ssstyle X\kern-.1em X}}

\let\ssstyle=\scriptscriptstyle

\def\Kout{\setbox1=\hbox{\Huge\bf K}\hbox to
1.05\wd1{\hspace{.05\wd1}
}}
\def\Sout{\setbox1=\hbox{\Huge\bf S}\hbox to 1.05\wd1{\hspace{.05\wd1}
}}

%% file: labelfig.tex
  \ifx\LabelFigloaded\MYundefined\relax
  \else
   \fi

  \def\LabelFigloaded{\relax}

  \chardef\LabelFigCatAt\the\catcode`\@
  \catcode`\@=11

 \let\LabelFigwlog@ld\wlog
 \def\wlog#1{\relax}

 \ifx\\\MYundefined@
    \let\\\relax
 \fi

  \def\ms@g{\immediate\write16}

 \def\N@wif{\csname newif\endcsname }
 \def\Temp@ {\N@wif\ifIN@}
 \ifx\INN@\MYundefined@
    \else \let\Temp@\relax
 \fi
 \Temp@

  \def\IN@{\expandafter\INN@\expandafter}
  \long\def\INN@0#1@#2@{\long\def\NI@##1#1##2##3\ENDNI@
    {\ifx\m@rker##2\IN@false\else\IN@true\fi}%
     \expandafter\NI@#2@@#1\m@rker\ENDNI@}
  \def\m@rker{\m@@rker}
 
  \newtoks\Initialtoks@  \newtoks\Terminaltoks@
  \def\SPLIT@{\expandafter\SPLITT@\expandafter}
  \def\SPLITT@0#1@#2@{\def\TTILPS@##1#1##2@{%
     \Initialtoks@{##1}\Terminaltoks@{##2}}\expandafter\TTILPS@#2@}

 \def\Shifted@@#1#2#3{\setbox0=\hbox{#3}%
   \raise -\dp0\vbox {\kern-#2%
       \hbox {\kern#1\unhbox0\kern-#1}%
           \kern#2}}

 \newcount\gridcount
 \newbox\auxGridbox@ \newbox\hGridbox@ \newbox\vGridbox@
 \newbox\Labelbox@ \newbox\auxLabelbox@
 \newbox\Coordinatebox@
 \newtoks\Labeltoks@
 \newdimen\Wdd@ \newdimen\Htt@
 \newdimen\Wddd@ \newdimen\Httt@
 
 \def\Wr@{\immediate\write16}

 \newdimen\GL@wd
 \def\GridLineWidth#1{\GL@wd=#1}

 \def\gobble#1{}
 \def\EdgeErr@{\Wr@{}%
      \Wr@{\string\Edges\space argument
      1, 10, 100 or 1000 please\string!}%
      }

 \newcount\Edgect@

 \def\Sweepup#1\endSweepup{}

 \def\SetEdges@{%
    \edef\Zr@@s{\expandafter\gobble\number\Edgect@\empty}%
        \count255=0\Zr@@s\relax
        \ifnum\count255=\z@\else\EdgeErr@\show\tailtest\fi
        \count255=1\Zr@@s\relax
        \ifnum\count255=\Edgect@\relax\else\EdgeErr@\show\leadtest\fi
    \EdgGl@b\edef\Zr@s{\expandafter\gobble\Zr@@s\empty}
    \ifnum\Edgect@>\@ne\relax\EdgGl@b\let\L@Dc\empty
        \else\EdgGl@b\edef\L@Dc{\string.}\fi
    \ifnum\Edgect@>\@ne\relax
        \EdgGl@b\edef\Edgescale@##1{\divide##1 by \Edgect@}%
        \else\EdgGl@b\edef\Edgescale@##1{}\fi
    }

 \def\Edges#1{\Edgect@=#1\relax
     \let\EdgGl@b\global \SetEdges@}

 \Edges{1}

 \def\hhrule{\hrule height \GL@wd\vskip-.\GL@wd}

 \def\hRule@{%
   \advance\gridcount -2%
   \vfil\hhrule\vfil
   \llap{\smash{\raise -2.5pt
     \hbox{\L@Dc\number\gridcount\Zr@s\kern2pt}}}%
   \hhrule
   }

\def\vvrule{\vrule width \GL@wd \kern-\GL@wd}

 \def\vRule@{\advance\gridcount 2%
   \hfil\vvrule\hfil
   \setbox\auxGridbox@=\vbox to 0pt
      {\vskip \Htt@\vskip 2pt
        \hbox to 0pt{\hss\L@Dc\number\gridcount\Zr@s\hss}\vss}%
      \wd\auxGridbox@=0pt \box\auxGridbox@
   \vvrule
   }

 \def\PlaceGrid@@{\gridcount=10 
  \setbox\hGridbox@=\hbox{%
        \hbox{%
             \hskip-.4pt\vrule
             \vbox to \Htt@{%
               \offinterlineskip\parindent=\z@\relax
               \hbox to \Wdd@{\hfil}
               \hRule@\hRule@\hRule@\hRule@
               \vfil\hhrule\vfil}%
             \vrule\hskip-.4pt}
    }%
  \gridcount=0%
  \setbox\vGridbox@=\hbox{%
      \vbox{\offinterlineskip\parindent=0pt\hsize=0pt
         \vskip-.4pt\hrule%
         \hbox to \Wdd@{%
                 \vtop to \Htt@{\vfil}%
                 \vRule@\vRule@\vRule@\vRule@
                 \hfil\vvrule\hfil}%
         \hrule\vskip-.4pt}}%
  \wd\hGridbox@=0pt\ht\hGridbox@=0pt
  \wd\vGridbox@=0pt\ht\vGridbox@=0pt
  \hbox{\box\hGridbox@\box\vGridbox@}%
  }

 \def\LabelsGlobal{\def\LabGl@b{\global}}
 \def\LabelsLocal{\def\LabGl@b{}}
 \LabelsGlobal 

 \def\SetLabels#1\endSetLabels{%
   \LabGl@b\Labeltoks@={#1()\\}%
   }

 \LabGl@b\Labeltoks@={()\\}

 \def\ShowGrid{\LabGl@b\let\PlaceGrid@\PlaceGrid@@}
 \def\HideGrid{\LabGl@b\let\PlaceGrid@\relax}
 \def\Grids{\ShowGrid\LabGl@b\let\GridSwitch@\ShowGrid}
 \def\noGrids{\HideGrid\LabGl@b\let\GridSwitch@\HideGrid}

 \noGrids

 \def\bAdjust@@{%
     \setbox\auxLabelbox@=\hbox{\raise \dp\auxLabelbox@
            \box\auxLabelbox@}}
 \def\bAdjust@{\let\vAdjust@\bAdjust@@}

 \def\eAdjust@@{\dimen0=-.5\ht\auxLabelbox@
     \advance\dimen0 by .5\dp\auxLabelbox@
     \setbox\auxLabelbox@=
            \hbox{\raise\dimen0\box\auxLabelbox@}}
 \def\eAdjust@{\let\vAdjust@\eAdjust@@}

 \def\tAdjust@@{%
     \setbox\auxLabelbox@=\hbox{\raise-\ht\auxLabelbox@
            \box\auxLabelbox@}}
 \def\tAdjust@{\let\vAdjust@\tAdjust@@}

 \let\vAdjust@\relax

 \def\lAdjust@{\let\hAdjust@\rlap}
 \def\rAdjust@{\let\hAdjust@\llap}

 \let\hAdjust@\relax\let\vAdjust@\relax

 \def\FetchLabel@#1(#2)#3\\{%
     \IN@0#2@@\ifIN@
        \setbox0=\hbox{\ignorespaces#1#3\unskip}%
        \ifdim\wd0>0pt
           \ms@g{}%
           \ms@g{ !!! Bad label(s)? !!!}%
           \message{ #1(#2)#3}%
        \fi
        \def\LabelMole@##1\endFetchLabel@{%
            \IN@0()\\@##1@%
            \ifIN@\def\Temp@{\FetchLabel@##1\endFetchLabel@}%
            \else\def\Temp@{}%
            \fi
            \Temp@
           }%
     \else
       \ignorespaces#1\unskip
       \setbox\auxLabelbox@=%
         \hbox to 0pt{\hss\ignorespaces\hAdjust@
          {\ignorespaces#3\unskip}\hss}%
       \vAdjust@
       \let\hAdjust@\relax\let\vAdjust@\relax
       \AugmentLabelBox@@{#2}%
       \ht\Labelbox@=0pt\dp\Labelbox@=0pt
       \let\LabelMole@\FetchLabel@%
     \fi\LabelMole@}

 \newtoks\XYSep@ 
 \def\SetXYSeparator#1{%
     \IN@0#1@@\ifIN@\XYSep@{*}%
     \else
     \XYSep@{#1}%
     \fi
     }

 \SetXYSeparator*

 \def\AugmentLabelBox@@#1{%
     \IN@0\the\XYSep@ @#1@\ifIN@
       \SPLIT@0\the\XYSep@ @#1@%
       \setbox\Labelbox@=\hbox to 0pt{%
         \unhbox\Labelbox@
         \Shifted@@{\the\Initialtoks@\Wddd@}%
         {\the\Terminaltoks@\Httt@}%
         {\box\auxLabelbox@}}%
     \else
         \ms@g{}%
         \ms@g{ !!! Bad insertion point. !!!}%
         \message{ (#1\ this point was rejected.)}%
     \fi
    }

 \def\FetchOption@#1[#2]#3\endFetchOption@{%
    \def\temp{#1}
    \ifx\temp\empty
       \Edgect@=#2\relax
       \let\EdgGl@b\relax
       \SetEdges@
       \Cleaner@#3%
    \fi}

 \def\Cleaner@#1[@]{\Labeltoks@{#1}}
     
 \def\PlaceLabels@@{\mathsurround=0pt
     \def\Cr@{\\}%
     \let\L\lAdjust@\let\R\rAdjust@
     \let\B\bAdjust@\let\E\eAdjust@\let\T\tAdjust@
     \expandafter\FetchOption@\the\Labeltoks@[@]\endFetchOption@
     \Wddd@=\Wdd@ \Edgescale@\Wddd@ 
     \Httt@=\Htt@ \Edgescale@\Httt@
     \expandafter\FetchLabel@\the\Labeltoks@\endFetchLabel@
     \box\Labelbox@
     }%

 \let \PlaceLabels@\PlaceLabels@@

 \def\AffixLabels#1{\setbox\Coordinatebox@=\hbox{#1}%
      \Wdd@=\wd\Coordinatebox@ \Htt@=\ht\Coordinatebox@
      \advance\Htt@ \dp\Coordinatebox@
      \hbox{\copy\Coordinatebox@\kern-\Wdd@ 
           \Shifted@@{0pt}{-\dp\Coordinatebox@}%
           {\PlaceLabels@\PlaceGrid@}%
           \kern\Wdd@}%
      \GridSwitch@ 
      \LabGl@b\Labeltoks@{()\\}%
      }
 
   \let\wlog\LabelFigwlog@ld   
   \catcode`\@=\LabelFigCatAt  

                                By

              Raymond S\'eroul <A18645@FRCCSC21.BITNET>
                                and 
              Laurent Siebenmann <lcs@topo.math.u-psud.fr>
    
              VERSIONS: July 1991, Oct 1991, Jan 1992, July 1992

INTRODUCTION

      This labelling package is intended for TeX users who
rely on non-TeX sources for for their graphics inserts.  It
provides means for adding TeX labels to such inserts with a
minimum of fuss. 

       For most labels, TeX users have in the past found it
reasonably convenient to rely on non-TeX sources. Typical
occasions when an inescapable need for TeX labels seemed to
arise are

 (a) when the graphics program lacks certain exotic or complex
mathematical symbols

 (b) when the very highest typographical quality is wanted for the
labels

 (c) when labels included with the graphics fail to print, 
 and you cannot figure out why (cf. boxedeps.doc).  The labels
 provided by labelfig.tex are 100

       Since this package first appeared, many users, who in the
past scarcely dreamed of using TeX labels, have come to use
nothing but.  So it is now appropriate to add

Intoxication Warning:  TeX labels may be addictive and expensive. 

     If you have a fast preview you may disagree, and even find
that this package provides an agreeable paste-up environment; see
extra applications at end.

     Note to publishers: It is possible and convenient to ultimately
export the TeX labels produced by labelfig.tex to become an integral
part of the EPS file. This is often desired by a publisher who typically
uses an "upmarket" graphics or page layout program, with which the
staff is skilled in perfecting figures.  See Appendix I for
a recipe.

     The authors are grateful to Patrick Ion of Math Reviews for
helpful comments and encouragement.

BASIC INSTRUCTIONS

    After reading in the macro file using

\input labelfig.tex
preview or proof your figure with a coordinate grid printed on
top, by typing the following:

    \ShowGrid  
    \AffixLabels{<the graphics insertion>}

Here <the graphics insertion> is what you would type to insert
the graphics object alone without the grid.  This must provide
for the space around it. For example <the graphics insertion>
might well be \BoxedEPSF{MyFigure scaled 700} using the
boxedeps.tex macro package (from same source); this provides a
TeX box containing the encapsulated PostScript insert specified by
the file MyFigure. \AffixLabels{...} provides the grid (supposing
\ShowGrid is present) and later, once you have specified labels
using the grid, it will "tack on" the labels.

     The grid is a sort of (usually elongated) checkerboard of
ten rows and ten columns and its (internal) partitions are by
default numbered  .1, ... ,.9  both horizontally (X-coordinate
running left to right) and vertically (Y-coordinate running bottom
to top).  Thus the points enclosed by the grid correspond to the
points of the unit square in the cartesian "X-Y" plane, the lower
left corner corresponding to the origin (0,0).  By extrapolation,
the full page corresponds to a larger rectangle in the plane.

     These coordinates serve to position labels as follows.
Before the \AffixLabels{...} command type label specifications:

  \SetLabels
   (<X-coordinate>*<Y-coordinate>) <first label> \\
   .
   .
   .
   (<X-coordinate>*<Y-coordinate>)  <last label> \\
  \endSetLabels

Each row specifies one label and is terminated by \\.  In each
row, the position indicator comes first; it is written as a
standard cartesian point except that the X- and Y- coordinates
are separated by * rather than a comma because TeX allows a
comma as decimal point. There are no dimension units to specify
as the unit is the grid itself.

     By default, this cartesian point specifies where the middle
of the baseline of the label will be located.  However if you precede
the point by \L [or \R] the left [or right] edge of the baseline will
be located there. Similarly you may also precede the point by \T, \E,
or \B to vertically align the top equator or bottom of the label box
at the specified point.  This gives nine standard positions of
the label with respect to the insertion point --- corresponding to
the eight principle points of the compas and the center

                     \L\T     \T      \R\T

                     \L\E     \E      \R\E

                     \L\B     \B      \R\B

But this neglects the default "baseline" level of TeX,
giving potentially three more positions

                     \L    <no tag>   \R

For text, the baseline level is often the preferred. Its relation to
the others is variable. It will often coincide with the bottom level,
as happens for "X".  But it is often distinct, as for "g", in which
case you have in all 12 distinct positions rather than 9.

     It is convenient to think of this specification of label
position as attaching the label by a thumb-tack to the coordinate
grid. There are up to twelve positions of the thumb-tack on the
label, while the position of the thumb-tack on the coordinate grid is
arbitrary.  Normally, one choses the position of the thumb-tack on
the label to be the one that is the closest to the item being
labeled.  There are good reasons for this "rule of thumb":

   (a)  It facilitates correct positioning at first try.

   (b)  If the scale of the figure must be altered after labels
have been affixed, the labels have a good chance of remaining well
positioned.

   (c)  The visible grid need not extend beyond the "bounding box"
for the figure, because the best preferred position is always
(at least almost) within the bounding box .

The second reason is particularly important. Indeed it often
happens that scale has to be altered after labelling begins, in
order to either provide space for the labels, or to adjust
proportions between the labels and the figure.  (The size of labels
is unaffected by scaling.)

     Here is an artificial but self-contained test which uses
TeX rules to make a graphics object.

TEST

    Do not skip this!

\input labelfig

 \def\FrameIt#1{\hbox{\vrule$\vcenter {\hrule\kern3pt%
             \hbox {\kern3pt #1\kern3pt}%
               \kern3pt\hrule}$\relax\vrule}}

 \def\Caption#1#2{\FrameIt{%
       \vtop {\hsize=#1\relax \parindent=0pt
         \leftskip=0pt \rightskip=0pt plus15pt
         \parfillskip=0pt
         \lineskip=1pt\baselineskip=0pt
         #2}}}

 \def\FirstQuadrant{\hbox to 100pt{\vrule\vbox to 100pt{%
        \hbox to 100pt{\hfil}\vfil\hrule}\hss}}

  \SetLabels
    \R(.5*.2) $\zeta\,\cdot$\\
    (.9*-.10) $\xi$\\
    \R(-.03*.9) $\eta$\\
    \T(.5*.9) \Caption{70pt}{%
          \it The norm of
          $g(\xi+i\eta)$ is indicated on
          contours of this invisible surface.}\\
  \endSetLabels

  \AffixLabels{\FirstQuadrant}

  \end

  Note that the coordinates to use for labels are indicated on the
edges of the grid (when visible) corresponding to the conventional
x- and y- axes of the Cartesian plane. By default the grid is
1-by-1. However, by the command \Edges{100}, you can change this
to 100-by-100 and many users find this alternative most
convenient. Place the command \Edges{...} in your style file (or
header) since its effect is is global. Other possible edge values
are 10 and 1000.

  If you use the command \Edges{...} at all, do so with care.  For
if you accidentally delete an \Edges{...} command your labels will
abruptly be badly misplaced and may logically but mysteriously
generate "dimension too big" errors under TeX and "off page" errors
under your driver.  

  You can dictate the edgescale for an individual figure by giving
the scale in brackets immediately after \SetLabels.  Thus, to
import into an article using say \Edge{100} a figure labelled using
another edgescale, say the original 1-by-1 default, you can use
\SetLabels[1]...\endSetLabels.

GETTING IT DOWN PAT

     Complicated labeling deserves the same respect as
complicated mathematics.  Do not expect it to come out perfect the
first time!  What is needed in either case is a mechanism to
repeatedly typeset troublesome pieces.

     One mechanism is always available.  One does complicated
labelling in a separate "test" file involving just the figure being
labelled;  a texpert will know how to \dump TeX's current state as
a temporary format that restarts rapidly at each retry.  Usually,
one then pastes the completed labelled figure back into the main
TeX file, but, of course, one can also \input it as an auxiliary
file.

     If you do not have a TeXpert at handy, here is a first
approximation to an efficient setup. By deletions reduce a copy
of your article to just a few lines before and after the figure.
Now label the figure, and finally, copy and paste the labelled
figure to the original article. Then copy the next figure to label
into this testbed and repeat. The TeXpert can improve the  speed
at which TeX starts up, by compiling a format specifically for
your article; just one caution: best NOT include in the format
ephemeral details of setup like \Set<mydriver>ArtSpecials (from
boxedeps.tex because this reads  figure dimensions which you may
change during your work session.

     An improved mechanism to repeatedly typeset troublesome
pieces is now available on the Macintosh; it is called LinoTeX;
see the same ftp sources.  It could be set up on many types
of computer.

     Before using labelfig.tex to attach labels to a graphics
object inserted using boxedeps.tex or BoxedArt.tex, make it a
firm rule to carefully adjust the bounding box using the trimming
commands of these packages, and also at least tentatively scale
and position the object. Beware of changing the grid inadvertently
after the labels have been positioned.  For example, correcting
the bounding box of a PostScript graphics object can foul up the
labels by changing the coordinate grid to which the labels are
attached. This is particularly true for the trimming  commands of
boxedeps.tex and BoxedArt.tex. However, as noted already, change
of scale is much less disruptive, and modest adjustments should be
well tolerated.

     Sometimes the labels protrude so far from the bounding box
of a figure that the figure has to be repositioned.  Best do this
by ad hoc spacing, say using \hglue and \vglue; altering the
bounding box would create a vicious circle.

     Remember that you are responsible for preventing labels
from overlapping. You are responsible for all label typography
including size and style. A label is really just about anything
that can be put in a TeX box. Note that spaces at the beginning
and end of labels will normally be suppressed; if you really want
them you must protect them with TeX braces.

     This package temporarily sets the \mathsurround parameter
of TeX to zero  while the labels are being affixed. This is done
because nonzero \mathsurround space would influence the position
of left and right aligned labels; then, when a texpert or printer
modifies mathsurround, diagram labeling might be disastrously
altered. There is a small price to pay involving labels that are
formatted as caption boxes including mathematics: you  may want or
need to specify an explicit mathsurround space within the caption
box; it will not influence anything outside.

     Those hostile to the use of * as separator between
the X and Y coordinates of label insertion points, are free to
impose another using \SetXYSeparator{<the new separator>}.  
Americans may prefer "," to "*" since they never use a 
comma as a decimal point; on the other hand, * may be more visible.

APPENDIX (I)  MERGING labelfig.tex LABELS INTO AN EPSF GRAPHICS OBJECT.

     As promised in the introduction, here is a recipe useful for
publishers. It works at least on Macintosh and at least for vectorized
graphics and Adobe type1 fonts.  (There is surely a similar recipe for
PCs under MSWindows.)

 (a)  Use boxedeps.tex utility to integrate the figure given by the eps
file, "x.eps" say, with a visible frame around it.  See
\ShowDisplacementBoxes command in boxedeps.tex.  To get precise results
automatically it is important to use the \Trim... commands of
boxedeps.tex making the "DisplacementBox" neatly fit the figure.

 (b)  Use the TeX printer driver and LaserWriter (versions >= 8.1.1) to
export to an EPSF the DVI page containing the integrated, labelled
figure. You now have an EPS file  "xx.eps"  that contains too much, and at
the wrong scale, and at wrong position.

 (c)  Convert the EPSF to an Adode Illustrator format EPSF using
the shareware utility called epsConvert by Sam Weiss
1993-- (currently $25).

 (d)  In Illustrator (or a compatible program), group the labels and the
"DisplacementBox"; copy them to the clipboard and paste them into "x.ps".
This step requires that all the label fonts be "visible to the Macintosh.

 (e)  Translate and scale the pasted group consisting of the labels plus
the "DisplacementBox" so as to make the "DisplacementBox" the bounding
box of (labelless) figure represented by "x.eps".  At this point the
labels will be correctly placed on the figure "x.eps".

 (f)  Ungroup and delete the "DisplacementBox".  The result is the
desired single EPS file, "x+.eps" say, It contains the original figure
plus its labels.  

     Using grouping and ungrouping appropriately in "x+.eps", a
publisher's staff can very efficiently improve label positions etc.

APPENDIX II)  SOME EXOTIC APPLICATIONS

     The grid of labelfig.tex is analogous to a light-table in
classical page makeup with wax or latex glue.  In principle, you
can use it to compose any page from its indivisible parts.  This
even has some of the artisanal charm of classical paste-up
provided you have a fast screen preview to make the process
"interactive".

     In practice labelfig.tex is a tool for nonstandard jobs.
Here are a few going beyond the labelling already discussed.

(I)  GRAPHICS INTEGRATION.

     This is accomplished by treating the imported graphics
objects as labels.  The underlying graphics object is then
typically an empty  \vbox to <dimension>{\vfill} in a TeX
\midinsert...\endinsert construction.  A label line
might be of the form

   (.1*.1) \\

The exact form of the special command varies from driver to
driver.  However, in the case of encapsulated PostScript graphics
(EPSF norm), by relying on boxedeps.tex, one can have the
following standard syntax (independant of driver  (see
boxedeps.doc for details.
  
  (.1*.1) \BoxedEPSF{MyFigure scaled <scale in mils>}\\

This may be slow since it requires TeX to read the PostScript
file to read bounding box using many complex macros.  So you
may want to try

  (.1*.1) \EPSFSpecial{MyFigure}{<scale in mils>}\\

which is fast and driver independant, but it squashes the
bounding box, normally to its lower left corner.

     Similarly for graphics of the Macintosh PICT norm ---
using BoxedArt.tex (same sources) in place of boxedeps.tex.

     This approach to integration is to be recommended when
one is assembling a composite graphics object.

 (II)  COMMUTATIVE DIAGRAM ENHANCEMENT

     Commutative diagrams or arrays of mathematical objects
connected by arrows of various sorts are common in mathematics.
The mathematical objects require the use of TeX.  Recently TeX
acquired a good collection of arrows of all slopes --- that of
LamSTeX --- plus pwerful macros to build the diagrams.

     However, even the LamSTeX collection is often
inadequate; it lacks for example double shafted arrows, dotted
arrows and curved arrows. Fortunately it is possible to produce
such arrows on an individual basis using sophisticated graphics
programs such as Illustrator and AldusFreehand (both serving
the EPSF norm) or using Metafont (with its public domain norm).
Since the creation of each new arrow is a work of love, you
probably want to limit the number of arrows by using LamSTeX
for most arrows. The 40K commutative diagram module of LamSTeX
has been adapted to work with AmSTeX and a copy may be posted
with LabelFig and related files. Unfortunately no one has yet
offered a version that works with Plain TeX or LaTeX.

       Suffice it here to say that when the exotic arrow has
been somehow imported into TeX, labelfig.tex treats it as a
label that one affixes to the commutative diagram.  Two other
steps will be treated in separate notes, namely the matter of
extracting the dimension specifications for the arrow and the
construction of the arrow --- for these steps are far from
unique and often depend intimately on your computer environment. 
Notes for the Macintosh-Textures-Illustrator combination are
found in the file ExoticArrows.doc.

 (III) NESTING 

Ingenuity pays off in exploiting labelfig.tex. One can
mix graphics and typography quite freely.  labelfig.tex is good
for freeform or overlapping arrangements, while boxedeps.tex (or
BoxedArt.tex) is best for regimented non-overlapping
arrangements --- and the two can be combined.

     The default behavior of labelfig.tex is not ideal 
for nesting objects, because to prevent trouble for beginners
the register for labels is globally cleared when \AffixLabels
concludes.  But there are switches available

      \LabelsGlobal      \LabelsLocal

which change this.  To understand this, extend the above test 
by something like:

 \LabelsLocal

 \SetLabels
    (.5*.5) AAA\\
 \endSetLabels

 {
 \SetLabels
    (.5*.5) ZZZ\\
 \endSetLabels
   \AffixLabels{\FirstQuadrant}
 }

   \AffixLabels{\FirstQuadrant}

     There are however potential pitfalls.  Neither
labelfig.tex nor boxedeps.tex has been tested under extreme
conditions. Problems may occur if their procedures are
indiscriminately nested. For boxedeps.tex (not labelfig.tex)
there is a precise cause for worry, namely many of its
variables are "global", which means that TeX braces will not
provide the protection one might expect.

COMMAND SUMMARY FOR labelfig.tex

  Here [...] means optional (one or zero)
       [...]* means any number of such constructs

  \SetLabels
    [[<P>](<X><Sep><Y>) <label> \\]*
  \endSetLabels
  \ShowGrid  
  \AffixLabels{<the figure>}

   --- <P> is tack position, one of eleven or empty
              order irrelevant

                   \L\T      \T      \R\T

                   \L\E      \E      \R\E

                     \L               \R

                   \L\B      \B      \R\B

   --- (<X><Sep><Y>) insertion point;
  <Sep> is separator, = * by default;
  \SetXYSeparator{<Sep>} changes it.
   <X> and <Y> are real numbers

  --- <label> a label to attach 

  --- <the figure> the figure to label 

  \GlobalLabels (default)     
  \LocalLabels  setting for nested constructs.

 \Grids makes ALL grids appear; \HideGrid then makes just next disappear.
 \noGrids returns to default.  The commands are always global.

 \GridLineWidth{<dimension>} adjusts width of grid lines. Default is very
small, to give "hairline" effect. If your grid lines are missing try
setting \GridLineWidth{1pt}.

 \Edges#1 globally changes the edge size of all grids to the numerical 
value #1, which must be 1, 10, 100, or 1000.  The default is 1.

VERSION HISTORY.
 --- Jan 1993: \Edges#1 and [??] option after \SetLabels
 --- July 1992: \Grids, \noGrids, \HideGrid;
       Gridlines become hairlines; \GridLineWidth{<dimension>}.
 --- Oct 1991, Jan 1992: \SetXYSeparator{<Sep>},  \LabelsGlobal,
       \LabelsLocal.
 --- July 1991: first release

Address for bugs and other feedback:

        Raymond S\'eroul
        IREM and Lab. de Typographie Informatise
        Univ. Rene Descartes
        Strasbourg

    Tel 33-88-41-63-45
    Email:  A18645@FRCCSC21.BITNET

        Laurent Siebenmann
        Mathematique, Bat. 425,
        Univ de Paris-Sud,
        91405-Orsay,
        France

    Tel 33-1-6941-7949; 
    Email: lcs@topo.math.u-psud.fr